ICRC
The Astroparticle Physics Conference
34th International Cosmic Ray Conference
July 30 - August 6, 2015
The Hague, The Netherlands

# Pierre Auger Observatory and Telescope Array: Joint Contributions to the 34th International Cosmic Ray Conference (ICRC 2015)

## The Telescope Array Collaboration

R.U. Abbasi[1], M. Abe[2], T. Abu-Zayyad[1], M. Allen[1], R. Azuma[3], E. Barcikowski[1], J.W. Belz[1], D.R. Bergman[1], S.A. Blake[1], R. Cady[1], M.J. Chae[4], B.G. Cheon[5], J. Chiba[6], M. Chikawa[7], W.R. Cho[8], T. Fujii[9], M. Fukushima[9,10], T. Goto[11], W. Hanlon[1], Y. Hayashi[11], N. Hayashida[12], K. Hibino[12], K. Honda[13], D. Ikeda[9], N. Inoue[2], T. Ishii[13], R. Ishimori[3], H. Ito[14], D. Ivanov[1], C.C.H. Jui[1], K. Kadota[15], F. Kakimoto[3], O. Kalashev[16], K. Kasahara[17], H. Kawai[18], S. Kawakami[11], S. Kawana[2], K. Kawata[9], E. Kido[9], H.B. Kim[5], J.H. Kim[1], J.H. Kim[19], S. Kitamura[3], Y. Kitamura[3], V. Kuzmin[16†], Y.J. Kwon[8], J. Lan[1], S.I. Lim[4], J.P. Lundquist[1], K. Machida[13], K. Martens[10], T. Matsuda[20], T. Matsuyama[11], J.N. Matthews[1], M. Minamino[11], Y. Mukai[13], I. Myers[1], K. Nagasawa[2], S. Nagataki[14], T. Nakamura[21], T. Nonaka[9], A. Nozato[7], S. Ogio[11], J. Ogura[3], M. Ohnishi[9], H. Ohoka[9], K. Oki[9], T. Okuda[22], M. Ono[23], A. Oshima[24], S. Ozawa[17], I.H. Park[25], M.S. Pshirkov[16,26], D.C. Rodriguez[1], G. Rubtsov[16], D. Ryu[19], H. Sagawa[9], N. Sakurai[11], L.M. Scott[27], P.D. Shah[1], F. Shibata[13], T. Shibata[9], H. Shimodaira[9], B.K. Shin[5], H.S. Shin[9], J.D. Smith[1], P. Sokolsky[1], R.W. Springer[1], B.T. Stokes[1], S.R. Stratton[1,27], T.A. Stroman[1], T. Suzawa[2], M. Takamura[6], M. Takeda[9], R. Takeishi[9], A. Taketa[28], M. Takita[9], Y. Tameda[12], H. Tanaka[11], K. Tanaka[29], M. Tanaka[20], S.B. Thomas[1], G.B. Thomson[1], P. Tinyakov[30,16], I. Tkachev[16], H. Tokuno[3], T. Tomida[31], S. Troitsky[16], Y. Tsunesada[3], K. Tsutsumi[3], Y. Uchihori[32], S. Udo[12], F. Urban[30], G. Vasiloff[1], T. Wong[1], R. Yamane[11], H. Yamaoka[20], K. Yamazaki[28], J. Yang[4], K. Yashiro[6], Y. Yoneda[11], S. Yoshida[18], H. Yoshii[33], R. Zollinger[1], Z. Zundel[1]

[1] High Energy Astrophysics Institute and Department of Physics and Astronomy, University of Utah, Salt Lake City, Utah, USA
[2] The Graduate School of Science and Engineering, Saitama University, Saitama, Saitama, Japan
[3] Graduate School of Science and Engineering, Tokyo Institute of Technology, Meguro, Tokyo, Japan
[4] Department of Physics and Institute for the Early Universe, Ewha Womans University, Seodaaemun-gu, Seoul, Korea
[5] Department of Physics and The Research Institute of Natural Science, Hanyang University, Seongdong-gu, Seoul, Korea
[6] Department of Physics, Tokyo University of Science, Noda, Chiba, Japan
[7] Department of Physics, Kinki University, Higashi Osaka, Osaka, Japan
[8] Department of Physics, Yonsei University, Seodaemun-gu, Seoul, Korea
[9] Institute for Cosmic Ray Research, University of Tokyo, Kashiwa, Chiba, Japan
[10] Kavli Institute for the Physics and Mathematics of the Universe (WPI), Todai Institutes for Advanced Study, the University of Tokyo, Kashiwa, Chiba, Japan
[11] Graduate School of Science, Osaka City University, Osaka, Osaka, Japan
[12] Faculty of Engineering, Kanagawa University, Yokohama, Kanagawa, Japan

[13] Interdisciplinary Graduate School of Medicine and Engineering, University of Yamanashi, Kofu, Yamanashi, Japan
[14] Astrophysical Big Bang Laboratory, RIKEN, Wako, Saitama, Japan
[15] Department of Physics, Tokyo City University, Setagaya-ku, Tokyo, Japan
[16] Institute for Nuclear Research of the Russian Academy of Sciences, Moscow, Russia
[17] Advanced Research Institute for Science and Engineering, Waseda University, Shinjuku-ku, Tokyo, Japan
[18] Department of Physics, Chiba University, Chiba, Chiba, Japan
[19] Department of Physics, School of Natural Sciences, Ulsan National Institute of Science and Technology, UNIST-gil, Ulsan, Korea
[20] Institute of Particle and Nuclear Studies, KEK, Tsukuba, Ibaraki, Japan
[21] Faculty of Science, Kochi University, Kochi, Kochi, Japan
[22] Department of Physical Sciences, Ritsumeikan University, Kusatsu, Shiga, Japan
[23] Department of Physics, Kyushu University, Fukuoka, Fukuoka, Japan
[24] Engineering Science Laboratory, Chubu University, Kasugai, Aichi, Japan
[25] Department of Physics, Sungkyunkwan University, Jang-an-gu, Suwon, Korea
[26] Sternberg Astronomical Institute, Moscow M.V. Lomonosov State University, Moscow, Russia
[27] Department of Physics and Astronomy, Rutgers University – The State University of New Jersey, Piscataway, New Jersey, USA
[28] Earthquake Research Institute, University of Tokyo, Bunkyo-ku, Tokyo, Japan
[29] Graduate School of Information Sciences, Hiroshima City University, Hiroshima, Hiroshima, Japan
[30] Service de Physique Théorique, Université Libre de Bruxelles, Brussels, Belgium
[31] Department of Computer Science and Engineering, Shinshu University, Nagano, Nagano, Japan
[32] National Institute of Radiological Science, Chiba, Chiba, Japan
[33] Department of Physics, Ehime University, Matsuyama, Ehime, Japan
[†] Deceased.

## The Pierre Auger Collaboration

A. Aab[41], P. Abreu[65], M. Aglietta[52], E.J. Ahn[80], I. Al Samarai[28], I.F.M. Albuquerque[16], I. Allekotte[1], P. Allison[85], A. Almela[11,8], J. Alvarez Castillo[58], J. Alvarez-Muñiz[75], R. Alves Batista[40], M. Ambrosio[43], A. Aminaei[59], G.A. Anastasi[45], L. Anchordoqui[79], S. Andringa[65], C. Aramo[43], F. Arqueros[72], N. Arsene[68], H. Asorey[1,24], P. Assis[65], J. Aublin[30], G. Avila[10], N. Awal[83], A.M. Badescu[69], C. Baus[35], J.J. Beatty[85], K.H. Becker[34], J.A. Bellido[12], C. Berat[31], M.E. Bertaina[52], X. Bertou[1], P.L. Biermann[38], P. Billoir[30], S.G. Blaess[12], A. Blanco[65], M. Blanco[30], J. Blazek[26], C. Bleve[47], H. Blümer[35,36], M. Boháčová[26], D. Boncioli[51], C. Bonifazi[22], N. Borodai[63], J. Brack[78], I. Brancus[66], T. Bretz[39], A. Bridgeman[36], P. Brogueira[65], P. Buchholz[41], A. Bueno[74], S. Buitink[59], M. Buscemi[43], K.S. Caballero-Mora[56], B. Caccianiga[42], L. Caccianiga[30], M. Candusso[44], L. Caramete[67], R. Caruso[45], A. Castellina[52], G. Cataldi[47], L. Cazon[65], R. Cester[46], A.G. Chavez[57], A. Chiavassa[52], J.A. Chinellato[17], J. Chudoba[26], M. Cilmo[43], R.W. Clay[12], G. Cocciolo[47], R. Colalillo[43], A. Coleman[86], L. Collica[42], M.R. Coluccia[47], R. Conceição[65], F. Contreras[9], M.J. Cooper[12], A. Cordier[29], S. Coutu[86], C.E. Covault[76], R. Dallier[33,32], B. Daniel[17], S. Dasso[5,3], K. Daumiller[36], B.R. Dawson[12], R.M. de Almeida[23], S.J. de Jong[59,61], G. De Mauro[59], J.R.T. de Mello Neto[22], I. De Mitri[47], J. de Oliveira[23], V. de Souza[15], L. del Peral[73], O. Deligny[28], N. Dhital[82], C. Di Giulio[44], A. Di Matteo[48], J.C. Diaz[82], M.L. Díaz Castro[17], F. Diogo[65], C. Dobrigkeit[17], W. Docters[60], J.C. D'Olivo[58], A. Dorofeev[78], Q. Dorosti Hasankiadeh[36], R.C. dos Anjos[15], M.T. Dova[4], J. Ebr[26], R. Engel[36], M. Erdmann[39], M. Erfani[41], C.O. Escobar[80,17], J. Espadanal[65], A. Etchegoyen[8,11], H. Falcke[59,62,61], K. Fang[87], G. Farrar[83], A.C. Fauth[17], N. Fazzini[80], A.P. Ferguson[76], B. Fick[82], J.M. Figueira[8], A. Filevich[8], A. Filipčič[70,71], O. Fratu[69], M.M. Freire[6], T. Fujii[87], B. García[7], D. García-Gámez[29], D. Garcia-Pinto[72], F. Gate[33], H. Gemmeke[37], A. Gherghel-Lascu[66], P.L. Ghia[30], U. Giaccari[22], M. Giammarchi[42], M. Giller[64], D. Głas[64], C. Glaser[39], H. Glass[80], G. Golup[1], M. Gómez Berisso[1], P.F. Gómez Vitale[10], N. González[8], B. Gookin[78], J. Gordon[85], A. Gorgi[52], P. Gorham[88], P. Gouffon[16], N. Griffith[85], A.F. Grillo[51], T.D. Grubb[12], F. Guarino[43], G.P. Guedes[18], M.R. Hampel[8], P. Hansen[4], D. Harari[1], T.A. Harrison[12], S. Hartmann[39], J.L. Harton[78], A. Haungs[36], T. Hebbeker[39], D. Heck[36], P. Heimann[41], A.E. Hervé[36], G.C. Hill[12], C. Hojvat[80], N. Hollon[87], E. Holt[36], P. Homola[34], J.R. Hörandel[59,61], P. Horvath[27], M. Hrabovský[27,26], D. Huber[35], T. Huege[36], A. Insolia[45], P.G. Isar[67], I. Jandt[34], S. Jansen[59,61], C. Jarne[4], J.A. Johnsen[77], M. Josebachuili[8], A. Kääpä[34], O. Kambeitz[35], K.H. Kampert[34], P. Kasper[80], I. Katkov[35], B. Keilhauer[36], E. Kemp[17], R.M. Kieckhafer[82], H.O. Klages[36], M. Kleifges[37], J. Kleinfeller[9], R. Krause[39], N. Krohm[34], D. Kuempel[39], G. Kukec Mezek[71], N. Kunka[37], A.W. Kuotb Awad[36], D. LaHurd[76], L. Latronico[52], R. Lauer[90], M. Lauscher[39], P. Lautridou[33], S. Le Coz[31], D. Lebrun[31], P. Lebrun[80], M.A. Leigui de Oliveira[21], A. Letessier-Selvon[30], I. Lhenry-Yvon[28], K. Link[35], L. Lopes[65], R. López[53], A. López Casado[75], K. Louedec[31], A. Lucero[8], M. Malacari[12], M. Mallamaci[42], J. Maller[33], D. Mandat[26], P. Mantsch[80], A.G. Mariazzi[4], V. Marin[33], I.C. Mariş[74], G. Marsella[47], D. Martello[47], H. Martinez[54], O. Martínez Bravo[53], D. Martraire[28], J.J. Masías Meza[3], H.J. Mathes[36], S. Mathys[34], J. Matthews[81], J.A.J. Matthews[90], G. Matthiae[44], D. Maurizio[13], E. Mayotte[77], P.O. Mazur[80], C. Medina[77], G. Medina-Tanco[58], R. Meissner[39], V.B.B. Mello[22], D. Melo[8], A. Menshikov[37], S. Messina[60], M.I. Micheletti[6], L. Middendorf[39], I.A. Minaya[72], L. Miramonti[42], B. Mitrica[66], L. Molina-Bueno[74], S. Mollerach[1], F. Montanet[31], C. Morello[52], M. Mostafá[86], C.A. Moura[21], G. Müller[39], M.A. Muller[17,20], S. Müller[36], S. Navas[74], P. Necesal[26], L. Nellen[58], A. Nelles[59,61], J. Neuser[34], P.H. Nguyen[12], M. Niculescu-Oglinzanu[66], M. Niechciol[41], L. Niemietz[34], T. Niggemann[39], D. Nitz[82], D. Nosek[25], V. Novotny[25], L. Nožka[27], L.A. Núñez[24], L. Ochilo[41], F. Oikonomou[86], A. Olinto[87], N. Pacheco[73], D. Pakk Selmi-Dei[17], M. Palatka[26], J. Pallotta[2], P. Papenbreer[34], G. Parente[75], A. Parra[53], T. Paul[79,84], M. Pech[26], J. Pękala[63], R. Pelayo[55], I.M. Pepe[19], L. Perrone[47], E. Petermann[89], C. Peters[39], S. Petrera[48,49], Y. Petrov[78], J. Phuntsok[86], R. Piegaia[3], T. Pierog[36], P. Pieroni[3], M. Pimenta[65], V. Pirronello[45], M. Platino[8], M. Plum[39], A. Porcelli[36], C. Porowski[63], R.R. Prado[15], P. Privitera[87], M. Prouza[26], E.J. Quel[2], S. Querchfeld[34], S. Quinn[76], J. Rautenberg[34], O. Ravel[33], D. Ravignani[8], D. Reinert[39], B. Revenu[33], J. Ridky[26], M. Risse[41], P. Ristori[2], V. Rizi[48], W. Rodrigues de Carvalho[75], J. Rodriguez

Rojo[9], M.D. Rodríguez-Frías[73], D. Rogozin[36], J. Rosado[72], M. Roth[36], E. Roulet[1], A.C. Rovero[5], S.J. Saffi[12], A. Saftoiu[66], F. Salamida[28,c], H. Salazar[53], A. Saleh[71], F. Salesa Greus[86], G. Salina[44], J.D. Sanabria Gomez[24], F. Sánchez[8], P. Sanchez-Lucas[74], E.M. Santos[16], E. Santos[17], F. Sarazin[77], B. Sarkar[34], R. Sarmento[65], C. Sarmiento-Cano[24], R. Sato[9], C. Scarso[9], M. Schauer[34], V. Scherini[47], H. Schieler[36], D. Schmidt[36], O. Scholten[60,b], H. Schoorlemmer[88], P. Schovánek[26], F.G. Schröder[36], A. Schulz[36], J. Schulz[59], J. Schumacher[39], S.J. Sciutto[4], A. Segreto[50], M. Settimo[30], A. Shadkam[81], R.C. Shellard[13], G. Sigl[40], O. Sima[68], A. Śmiałkowski[64], R. Šmída[36], G.R. Snow[89], P. Sommers[86], S. Sonntag[41], J. Sorokin[12], R. Squartini[9], Y.N. Srivastava[84], D. Stanca[66], S. Stanič[71], J. Stapleton[85], J. Stasielak[63], M. Stephan[39], A. Stutz[31], F. Suarez[8,11], M. Suarez Durán[24], T. Suomijärvi[28], A.D. Supanitsky[5], M.S. Sutherland[85], J. Swain[84], Z. Szadkowski[64], O.A. Taborda[1], A. Tapia[8], A. Tepe[41], V.M. Theodoro[17], O. Tibolla[56], C. Timmermans[59,61], C.J. Todero Peixoto[14], G. Toma[66], L. Tomankova[36], B. Tomé[65], A. Tonachini[46], G. Torralba Elipe[75], D. Torres Machado[22], P. Travnicek[26], M. Trini[71], R. Ulrich[36], M. Unger[83,36], M. Urban[39], J.F. Valdés Galicia[58], I. Valiño[75], L. Valore[43], G. van Aar[59], P. van Bodegom[12], A.M. van den Berg[60], S. van Velzen[59], A. van Vliet[40], E. Varela[53], B. Vargas Cárdenas[58], G. Varner[88], R. Vasquez[22], J.R. Vázquez[72], R.A. Vázquez[75], D. Veberič[36], V. Verzi[44], J. Vicha[26], M. Videla[8], L. Villaseñor[57], B. Vlcek[73], S. Vorobiov[71], H. Wahlberg[4], O. Wainberg[8,11], D. Walz[39], A.A. Watson[a], M. Weber[37], K. Weidenhaupt[39], A. Weindl[36], C. Welling[39], F. Werner[35], A. Widom[84], L. Wiencke[77], H. Wilczyński[63], T. Winchen[34], D. Wittkowski[34], B. Wundheiler[8], S. Wykes[59], L. Yang[71], T. Yapici[82], A. Yushkov[41], E. Zas[75], D. Zavrtanik[71,70], M. Zavrtanik[70,71], A. Zepeda[54], B. Zimmermann[37], M. Ziolkowski[41], F. Zuccarello[45]

[1] Centro Atómico Bariloche and Instituto Balseiro (CNEA-UNCuyo-CONICET), San Carlos de Bariloche, Argentina
[2] Centro de Investigaciones en Láseres y Aplicaciones, CITEDEF and CONICET, Villa Martelli, Argentina
[3] Departamento de Física, FCEyN, Universidad de Buenos Aires and CONICET, Buenos Aires, Argentina
[4] IFLP, Universidad Nacional de La Plata and CONICET, La Plata, Argentina
[5] Instituto de Astronomía y Física del Espacio (IAFE, CONICET-UBA), Buenos Aires, Argentina
[6] Instituto de Física de Rosario (IFIR) – CONICET/U.N.R. and Facultad de Ciencias Bioquímicas y Farmacéuticas U.N.R., Rosario, Argentina
[7] Instituto de Tecnologías en Detección y Astropartículas (CNEA, CONICET, UNSAM), and Universidad Tecnológica Nacional – Facultad Regional Mendoza (CONICET/CNEA), Mendoza, Argentina
[8] Instituto de Tecnologías en Detección y Astropartículas (CNEA, CONICET, UNSAM), Buenos Aires, Argentina
[9] Observatorio Pierre Auger, Malargüe, Argentina
[10] Observatorio Pierre Auger and Comisión Nacional de Energía Atómica, Malargüe, Argentina
[11] Universidad Tecnológica Nacional – Facultad Regional Buenos Aires, Buenos Aires, Argentina
[12] University of Adelaide, Adelaide, S.A., Australia
[13] Centro Brasileiro de Pesquisas Fisicas, Rio de Janeiro, RJ, Brazil
[14] Universidade de São Paulo, Escola de Engenharia de Lorena, Lorena, SP, Brazil
[15] Universidade de São Paulo, Instituto de Física de São Carlos, São Carlos, SP, Brazil
[16] Universidade de São Paulo, Instituto de Física, São Paulo, SP, Brazil
[17] Universidade Estadual de Campinas, IFGW, Campinas, SP, Brazil
[18] Universidade Estadual de Feira de Santana, Feira de Santana, Brazil
[19] Universidade Federal da Bahia, Salvador, BA, Brazil
[20] Universidade Federal de Pelotas, Pelotas, RS, Brazil
[21] Universidade Federal do ABC, Santo André, SP, Brazil
[22] Universidade Federal do Rio de Janeiro, Instituto de Física, Rio de Janeiro, RJ, Brazil

[23] Universidade Federal Fluminense, EEIMVR, Volta Redonda, RJ, Brazil
[24] Universidad Industrial de Santander, Bucaramanga, Colombia
[25] Charles University, Faculty of Mathematics and Physics, Institute of Particle and Nuclear Physics, Prague, Czech Republic
[26] Institute of Physics of the Academy of Sciences of the Czech Republic, Prague, Czech Republic
[27] Palacky University, RCPTM, Olomouc, Czech Republic
[28] Institut de Physique Nucléaire d'Orsay (IPNO), Université Paris 11, CNRS-IN2P3, Orsay, France
[29] Laboratoire de l'Accélérateur Linéaire (LAL), Université Paris 11, CNRS-IN2P3, Orsay, France
[30] Laboratoire de Physique Nucléaire et de Hautes Energies (LPNHE), Universités Paris 6 et Paris 7, CNRS-IN2P3, Paris, France
[31] Laboratoire de Physique Subatomique et de Cosmologie (LPSC), Université Grenoble-Alpes, CNRS/IN2P3, Grenoble, France
[32] Station de Radioastronomie de Nançay, Observatoire de Paris, CNRS/INSU, Nançay, France
[33] SUBATECH, École des Mines de Nantes, CNRS-IN2P3, Université de Nantes, Nantes, France
[34] Bergische Universität Wuppertal, Fachbereich C – Physik, Wuppertal, Germany
[35] Karlsruhe Institute of Technology – Campus South – Institut für Experimentelle Kernphysik (IEKP), Karlsruhe, Germany
[36] Karlsruhe Institute of Technology – Campus North – Institut für Kernphysik, Karlsruhe, Germany
[37] Karlsruhe Institute of Technology – Campus North – Institut für Prozessdatenverarbeitung und Elektronik, Karlsruhe, Germany
[38] Max-Planck-Institut für Radioastronomie, Bonn, Germany
[39] RWTH Aachen University, III. Physikalisches Institut A, Aachen, Germany
[40] Universität Hamburg, II. Institut für Theoretische Physik, Hamburg, Germany
[41] Universität Siegen, Fachbereich 7 Physik – Experimentelle Teilchenphysik, Siegen, Germany
[42] Università di Milano and Sezione INFN, Milan, Italy
[43] Università di Napoli "Federico II" and Sezione INFN, Napoli, Italy
[44] Università di Roma II "Tor Vergata" and Sezione INFN, Roma, Italy
[45] Università di Catania and Sezione INFN, Catania, Italy
[46] Università di Torino and Sezione INFN, Torino, Italy
[47] Dipartimento di Matematica e Fisica "E. De Giorgi" dell'Università del Salento and Sezione INFN, Lecce, Italy
[48] Dipartimento di Scienze Fisiche e Chimiche dell'Università dell'Aquila and Sezione INFN, L'Aquila, Italy
[49] Gran Sasso Science Institute (INFN), L'Aquila, Italy
[50] Istituto di Astrofisica Spaziale e Fisica Cosmica di Palermo (INAF), Palermo, Italy
[51] INFN, Laboratori Nazionali del Gran Sasso, Assergi (L'Aquila), Italy
[52] Osservatorio Astrofisico di Torino (INAF), Università di Torino and Sezione INFN, Torino, Italy
[53] Benemérita Universidad Autónoma de Puebla, Puebla, México
[54] Centro de Investigación y de Estudios Avanzados del IPN (CINVESTAV), México, D.F., México
[55] Unidad Profesional Interdisciplinaria en Ingeniería y Tecnologías Avanzadas del Instituto Politécnico Nacional (UPIITA-IPN), México, D.F., México
[56] Universidad Autónoma de Chiapas, Tuxtla Gutiérrez, Chiapas, México
[57] Universidad Michoacana de San Nicolás de Hidalgo, Morelia, Michoacán, México
[58] Universidad Nacional Autónoma de México, México, D.F., México
[59] IMAPP, Radboud University Nijmegen, Nijmegen, Netherlands
[60] KVI – Center for Advanced Radiation Technology, University of Groningen, Groningen, Netherlands
[61] Nikhef, Science Park, Amsterdam, Netherlands
[62] ASTRON, Dwingeloo, Netherlands
[63] Institute of Nuclear Physics PAN, Krakow, Poland
[64] University of Łódź, Łódź, Poland
[65] Laboratório de Instrumentação e Física Experimental de Partículas (LIP) and Instituto Superior Técnico, Universidade de Lisboa (UL), Portugal

[66] "Horia Hulubei" National Institute for Physics and Nuclear Engineering, Bucharest-Magurele, Romania
[67] Institute of Space Science, Bucharest-Magurele, Romania
[68] University of Bucharest, Physics Department, Bucharest, Romania
[69] University Politehnica of Bucharest, Bucharest, Romania
[70] Experimental Particle Physics Department, J. Stefan Institute, Ljubljana, Slovenia
[71] Laboratory for Astroparticle Physics, University of Nova Gorica, Nova Gorica, Slovenia
[72] Universidad Complutense de Madrid, Madrid, Spain
[73] Universidad de Alcalá, Alcalá de Henares, Madrid, Spain
[74] Universidad de Granada and C.A.F.P.E., Granada, Spain
[75] Universidad de Santiago de Compostela, Santiago de Compostela, Spain
[76] Case Western Reserve University, Cleveland, OH, USA
[77] Colorado School of Mines, Golden, CO, USA
[78] Colorado State University, Fort Collins, CO, USA
[79] Department of Physics and Astronomy, Lehman College, City University of New York, Bronx, NY, USA
[80] Fermilab, Batavia, IL, USA
[81] Louisiana State University, Baton Rouge, LA, USA
[82] Michigan Technological University, Houghton, MI, USA
[83] New York University, New York, NY, USA
[84] Northeastern University, Boston, MA, USA
[85] Ohio State University, Columbus, OH, USA
[86] Pennsylvania State University, University Park, PA, USA
[87] University of Chicago, Enrico Fermi Institute, Chicago, IL, USA
[88] University of Hawaii, Honolulu, HI, USA
[89] University of Nebraska, Lincoln, NE, USA
[90] University of New Mexico, Albuquerque, NM, USA
[a] School of Physics and Astronomy, University of Leeds, Leeds, United Kingdom
[b] Also at Vrije Universiteit Brussels, Brussels, Belgium
[c] Currently at INFN Milano Bicocca, Milano, Italy

## Acknowledgments of the Telescope Array Collaboration

The Telescope Array experiment is supported by the Japan Society for the Promotion of Science through Grants-in-Aid for Scientific Research on Specially Promoted Research (21000002) "Extreme Phenomena in the Universe Explored by Highest Energy Cosmic Rays" and for Scientific Research (19104006), and the Inter-University Research Program of the Institute for Cosmic Ray Research; by the U.S. National Science Foundation awards PHY-0307098, PHY-0601915, PHY-0649681, PHY-0703893, PHY-0758342, PHY-0848320, PHY-1069280, PHY-1069286, PHY-1404495 and PHY-1404502; by the National Research Foundation of Korea (2007-0093860, R32-10130, 2012R1A1A2008381, 2013004883); by the Russian Academy of Sciences, RFBR grants 11-02-01528a and 13-02-01311a (INR), IISN project No. 4.4502.13, and Belgian Science Policy under IUAP VII/37 (ULB). The foundations of Dr. Ezekiel R. and Edna Wattis Dumke, Willard L. Eccles, and George S. and Dolores Doré Eccles all helped with generous donations. The State of Utah supported the project through its Economic Development Board, and the University of Utah through the Office of the Vice President for Research. The experimental site became available through the cooperation of the Utah School and Institutional Trust Lands Administration (SITLA), U.S. Bureau of Land Management, and the U.S. Air Force. We also wish to thank the people and the officials of Millard County, Utah for their steadfast and warm support. We gratefully acknowledge the contributions from the technical staffs of our home institutions. An allocation of computer time from the Center for High Performance Computing at the University of Utah is gratefully acknowledged.

## Acknowledgments of the Pierre Auger Collaboration

The successful installation, commissioning, and operation of the Pierre Auger Observatory would not have been possible without the strong commitment and effort from the technical and administrative staff in Malargüe. We are very grateful to the following agencies and organizations for financial support:

Comisión Nacional de Energía Atómica, Agencia Nacional de Promoción Científica y Tecnológica (ANPCyT), Consejo Nacional de Investigaciones Científicas y Técnicas (CONICET), Gobierno de la Provincia de Mendoza, Municipalidad de Malargüe, NDM Holdings and Valle Las Leñas, in gratitude for their continuing cooperation over land access, Argentina; the Australian Research Council; Conselho Nacional de Desenvolvimento Científico e Tecnológico (CNPq), Financiadora de Estudos e Projetos (FINEP), Fundação de Amparo à Pesquisa do Estado de Rio de Janeiro (FAPERJ), São Paulo Research Foundation (FAPESP) Grants No. 2010/07359-6 and No. 1999/05404-3, Ministério de Ciência e Tecnologia (MCT), Brazil; Grant No. MSMT-CR LG13007, No. 7AMB14AR005, and the Czech Science Foundation Grant No. 14-17501S, Czech Republic; Centre de Calcul IN2P3/CNRS, Centre National de la Recherche Scientifique (CNRS), Conseil Régional Ile-de-France, Département Physique Nucléaire et Corpusculaire (PNC-IN2P3/CNRS), Département Sciences de l'Univers (SDU-INSU/CNRS), Institut Lagrange de Paris (ILP) Grant No. LABEX ANR-10-LABX-63, within the Investissements d'Avenir Programme Grant No. ANR-11-IDEX-0004-02, France; Bundesministerium für Bildung und Forschung (BMBF), Deutsche Forschungsgemeinschaft (DFG), Finanzministerium Baden-Württemberg, Helmholtz Alliance for Astroparticle Physics (HAP), Helmholtz-Gemeinschaft Deutscher Forschungszentren (HGF), Ministerium für Wissenschaft und Forschung, Nordrhein Westfalen, Ministerium für Wissenschaft, Forschung und Kunst, Baden-Württemberg, Germany; Istituto Nazionale di Fisica Nucleare (INFN), Istituto Nazionale di Astrofisica (INAF), Ministero dell'Istruzione, dell'Universitá e della Ricerca (MIUR), Gran Sasso Center for Astroparticle Physics (CFA), CETEMPS Center of Excellence, Ministero degli Affari Esteri (MAE), Italy; Consejo Nacional de Ciencia y Tecnología (CONACYT), Mexico; Ministerie van Onderwijs, Cultuur en Wetenschap, Nederlandse Organisatie voor Wetenschappelijk Onderzoek (NWO), Stichting voor Fundamenteel Onderzoek der Materie (FOM), Netherlands; National Centre for Research and Development, Grants No. ERA-NET-ASPERA/01/11 and No. ERA-NET-ASPERA/02/11,

National Science Centre, Grants No. 2013/08/M/ST9/00322, No. 2013/08/M/ST9/00728 and No. HARMONIA 5 - 2013/10/M/ST9/00062, Poland; Portuguese national funds and FEDER funds within Programa Operacional Factores de Competitividade through Fundação para a Ciência e a Tecnologia (COMPETE), Portugal; Romanian Authority for Scientific Research ANCS, CNDI-UEFISCDI partnership projects Grants No. 20/2012 and No. 194/2012, Grants No. 1/ASPERA2/2012 ERA-NET, No. PN-II-RU-PD-2011-3-0145-17 and No. PN-II-RU-PD-2011-3-0062, the Minister of National Education, Programme Space Technology and Advanced Research (STAR), Grant No. 83/2013, Romania; Slovenian Research Agency, Slovenia; Comunidad de Madrid, FEDER funds, Ministerio de Educación y Ciencia, Xunta de Galicia, European Community 7th Framework Program, Grant No. FP7-PEOPLE-2012-IEF-328826, Spain; Science and Technology Facilities Council, United Kingdom; Department of Energy, Contracts No. DE-AC02-07CH11359, No. DE-FR02-04ER41300, No. DE-FG02-99ER41107 and No. DE-SC0011689, National Science Foundation, Grant No. 0450696, The Grainger Foundation, USA; NAFOSTED, Vietnam; Marie Curie-IRSES/EPLANET, European Particle Physics Latin American Network, European Union 7th Framework Program, Grant No. PIRSES-2009-GA-246806 and PIOF-GA-2013-624803; and UNESCO.

# Contributions

# Report of the Working Group on the Composition of Ultra-High Energy Cosmic Rays

**Michael Unger**[∗a,b], **for the Pierre Auger Collaboration**[c] **and the Telescope Array Collaboration**[d]

[a]*Center for Cosmology and Particle Physics, New York University, 4 Washington Place, New York, NY 10003, USA*
[b]*Institut für Kernphysik, Karlsruher Institut für Technologie, Postfach 3640, 76021 Karlsruhe, Germany*
[c]*Observatorio Pierre Auger, Av. San Martín Norte 304, 5613 Malargüe, Argentina*
*E-mail:* auger_spokespersons@fnal.gov
*Full author list:* http://www.auger.org/archive/authors_2015_ICRC.html
[d]*Telescope Array Project 201 James Fletcher Bldg. 115 S. 1400 Salt Lake City, UT 84112-0830*
*E-mail:* ta-icrc@cosmic.utah.edu
*Full author list:* http://www.telescopearray.org/images/papers/ICRC2015-authorlist.pdf

The atmospheric depth, $X_{\max}$, at which the particle number of an air shower reaches its maximum is a good indicator for the mass of the primary particle. We present a comparison of the energy evolution of the mean of $X_{\max}$ as measured by the Telescope Array and Pierre Auger Collaborations. After accounting for the different resolutions, acceptances and analysis strategies of the two experiments, the two results are found to be in good agreement within systematic uncertainties.

∗Speaker.

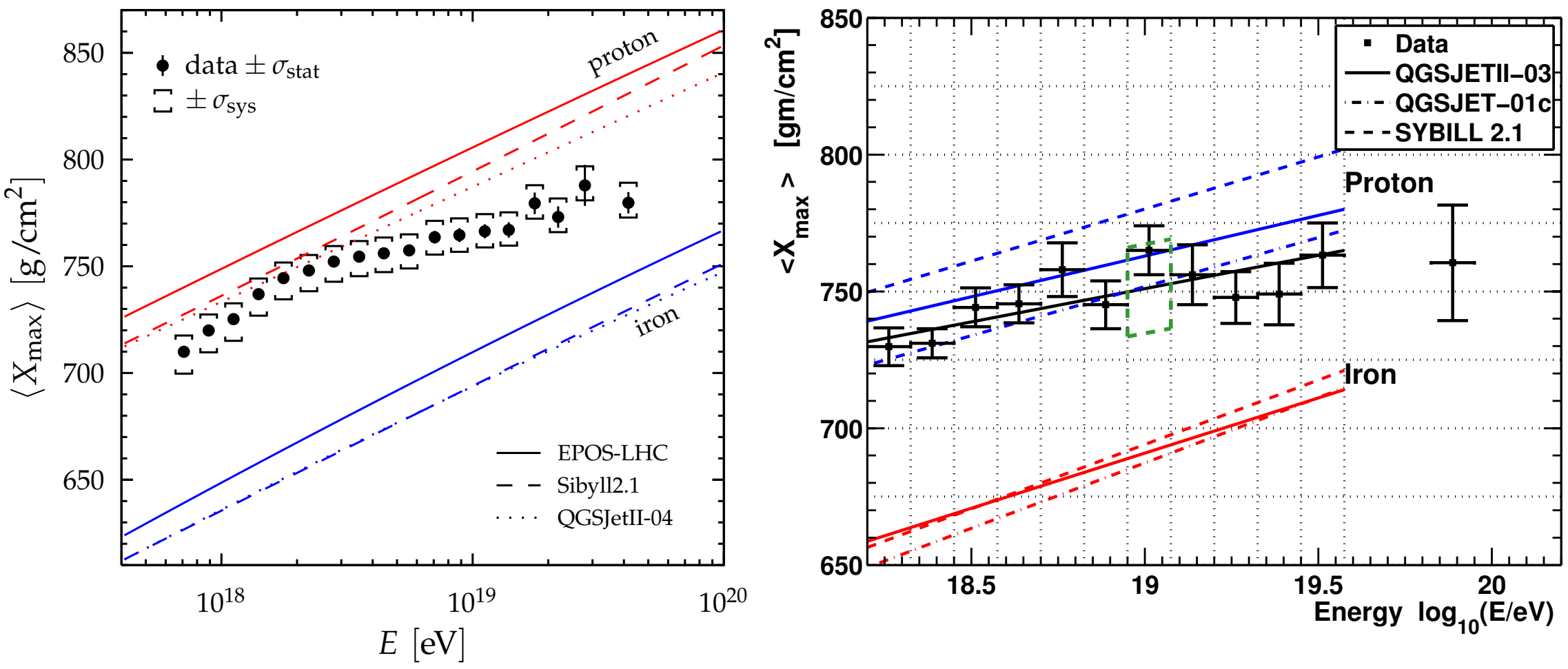

**Figure 1:** $\langle X_{\text{max}} \rangle$ as measured by the Pierre Auger (left) and Telescope Array (right) Collaborations [2, 3]. The colored lines denote predictions of air-shower simulations (note that different models are shown in the left and right panel, only SIBYLL2.1 is the same). The black line on the right panel is a straight-line fit to the TA data. Systematic uncertainties are indicated by brackets (left) and by the green dashed box (right).

## 1. Introduction

The nuclear composition of ultra-high energy cosmic rays is one of the key observables to understand their origin. One of the most robust and precise observables to date to infer the composition from air-shower measurements is the atmospheric depth at which the particle number of the shower reaches its maximum, $X_{\text{max}}$. Currently, the Pierre Auger Observatory and the Telescope Array (TA) measure $X_{\text{max}}$ using fluorescence detectors. But despite the use of the same detection principle, a direct comparison of the data published by both collaborations is not straightforward.

The TA Collaboration published values of the average shower maximum, $\langle X_{\text{max}} \rangle$, obtained from $X_{\text{max}}$ distributions that include detector effects such as the selection efficiency and acceptance. The interpretation of the data is made possible by the comparison of the Monte-Carlo prediction for proton and iron nuclei folded with the same detector resolution and efficiency. In the analysis performed by the Pierre Auger Collaboration, only shower geometries are selected allowing the sampling of almost unbiased $X_{\text{max}}$ distributions and residual biases from the acceptance, reconstruction and resolution are corrected for.

The corresponding values of $\langle X_{\text{max}} \rangle$ are presented in Fig. 1 together with predictions from air-shower simulations for proton- and iron-initiated showers. SIBYLL2.1, the only hadronic interaction model used by both collaborations, provides a common reference in these plots.

The work reported here is a common effort of the Auger and TA Collaborations with the aim of providing a direct comparison of the $\langle X_{\text{max}} \rangle$ measurements taking into account the different approaches of each collaboration. Indirect comparisons of TA and Auger results using a conversion of $\langle X_{\text{max}} \rangle$ to the average logarithmic mass were published in earlier [1]. The disadvantage of indirect comparisons is that they depend on the particular hadronic interaction model that is used. The current analysis was performed in the following way. The Auger $X_{\text{max}}$ distributions were fitted by a combination of four primary nuclei (proton, helium, nitrogen, iron) using events from air-shower simulations. The abundances which best fit the Auger data were simulated through the Middle Drum detector of TA (TA-MD) and analyzed by the TA Collaboration using the same procedure as applied to their data. This procedure resulted in the Auger data folded into the TA-MD detector. The Auger $\langle X_{\max}\rangle$ folded with TA-MD analysis is shown in this paper in comparison to the TA-MD data as published [3].

## 2. Data Samples

The analysis presented here is based on the data measured with the Pierre Auger Observatory in the period 1st December 2004 to 31st December 2012. All measured events were analyzed as explained in reference [2]. The events were selected to guarantee good measurement conditions and a high-quality reconstruction. After that, the fiducial selection was applied. In total 19,759 events were considered for further analysis (7365 above the lower energy threshold of TA, see below). The $X_{\max}$ values of these events were sampled in 18 energy bins starting at $\log(E/\mathrm{eV}) = 17.8$.

From the Telescope Array we use hybrid data collected with the MD fluorescence telescope and surface detector array over the period from the 27th May 2008 to 2nd May 2013. The reconstruction and analysis applied to the data are described in [3]. The number of events which passed all cuts is 438, for which the mean $X_{\max}$ is shown in 12 energy bins above $\log(E/\mathrm{eV}) = 18.2$.

The number of events used for this comparison presented here is shown in Fig. 2 and the $X_{\max}$-resolution of the two experiments is presented in Fig. 3. As can be seen, the resolutions after cuts are comparable but it is worthwhile noting that the resolution quoted for the MD does not contain effects from the detector calibration and atmospheric monitoring. The systematic uncertainties on the $X_{\max}$ scale, compared in the right panel of Fig. 3, are $\leq 10\,\mathrm{g/cm^2}$ and $16\,\mathrm{g/cm^2}$ for the Auger and TA analyses respectively.

## 3. Analysis

The relation between the true and observed $X_{\max}$ distribution is

$$f_{\mathrm{obs}}(X_{\max}^{\mathrm{rec}}) = \int_0^\infty f_{\mathrm{true}}(X_{\max})\,\varepsilon(X_{\max})\,R(X_{\max}^{\mathrm{rec}} - X_{\max})\,\mathrm{d}X_{\max}, \tag{3.1}$$

i.e., the true distribution $f_{\mathrm{true}}$ is deformed by the detection efficiency $\varepsilon$ and smeared by the detector resolution $R$ that relates the true $X_{\max}$ to the reconstructed one, $X_{\max}^{\mathrm{rec}}$.

Due to the different analysis approaches of the TA and Pierre Auger Collaborations it is not possible to compare the published values of the moments of the $X_{\max}$ distribution directly. Whereas $\langle X_{\max}\rangle$ and $\sigma(X_{\max})$ published by the Pierre Auger Collaboration are close to the true moments (i.e. the moments of $f_{\mathrm{true}}$), the TA collaboration published the $\langle X_{\max}\rangle$ folded with the effects of the detector response and reconstruction (i.e. the moments of $f_{\mathrm{obs}}$).

To be able to perform a comparison of the two results, we need to establish what $\langle X_{\max}\rangle_{\mathrm{obs}}$ would be if the $X_{\max}$ distributions measured by Auger were observed by the TA detector. For this purpose, we convolve a parametric description of $f_{\mathrm{true}}$ that is based on the Auger data with the TA detector simulation and apply the same reconstruction and analysis chain used for the TA data to this simulated data set (see [4] for a previous description of this method).

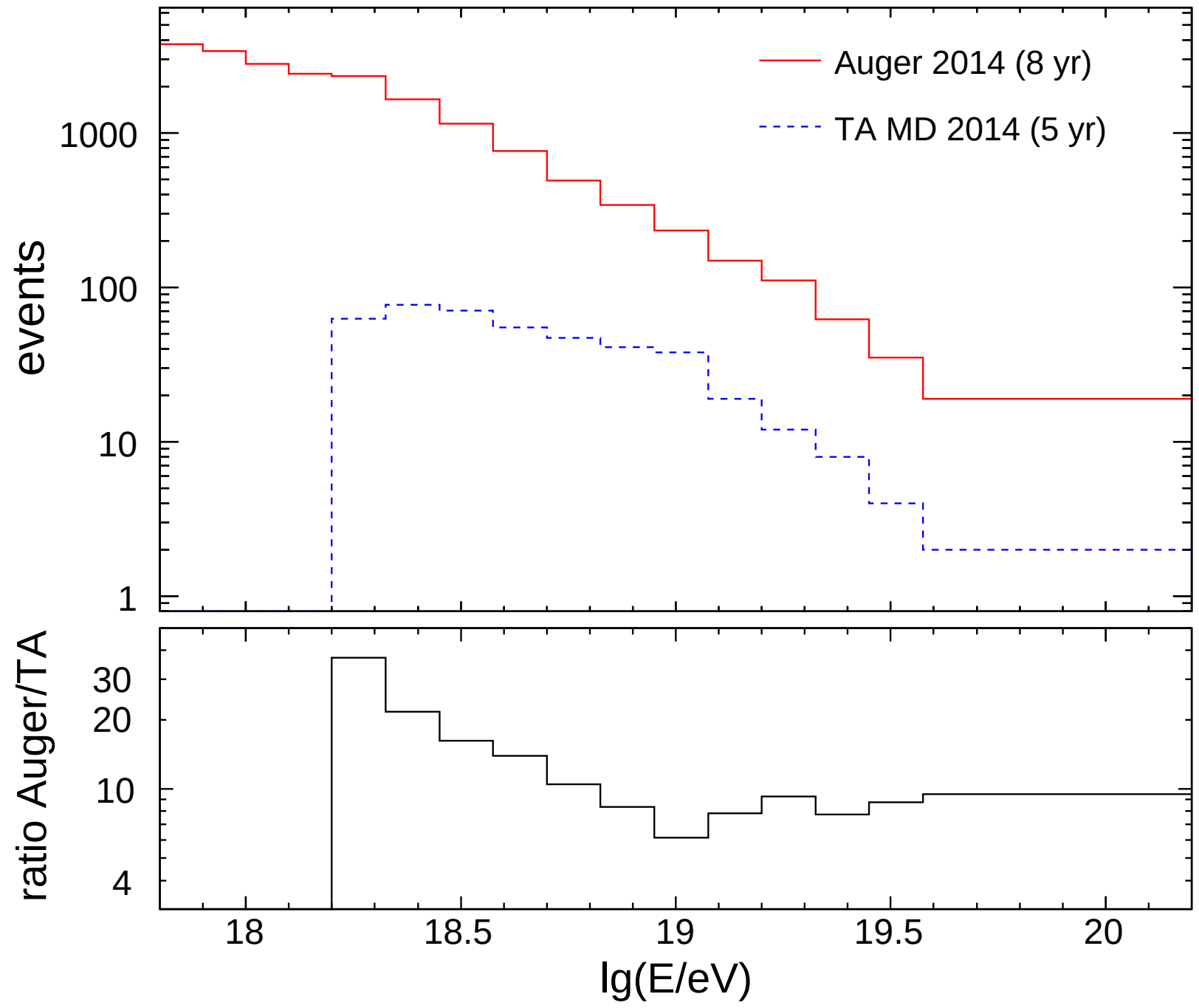

**Figure 2:** Number of selected events for the Auger (solid red line) and TA (blue dashed line) analyses. The ratio of events is given in the lower panel.

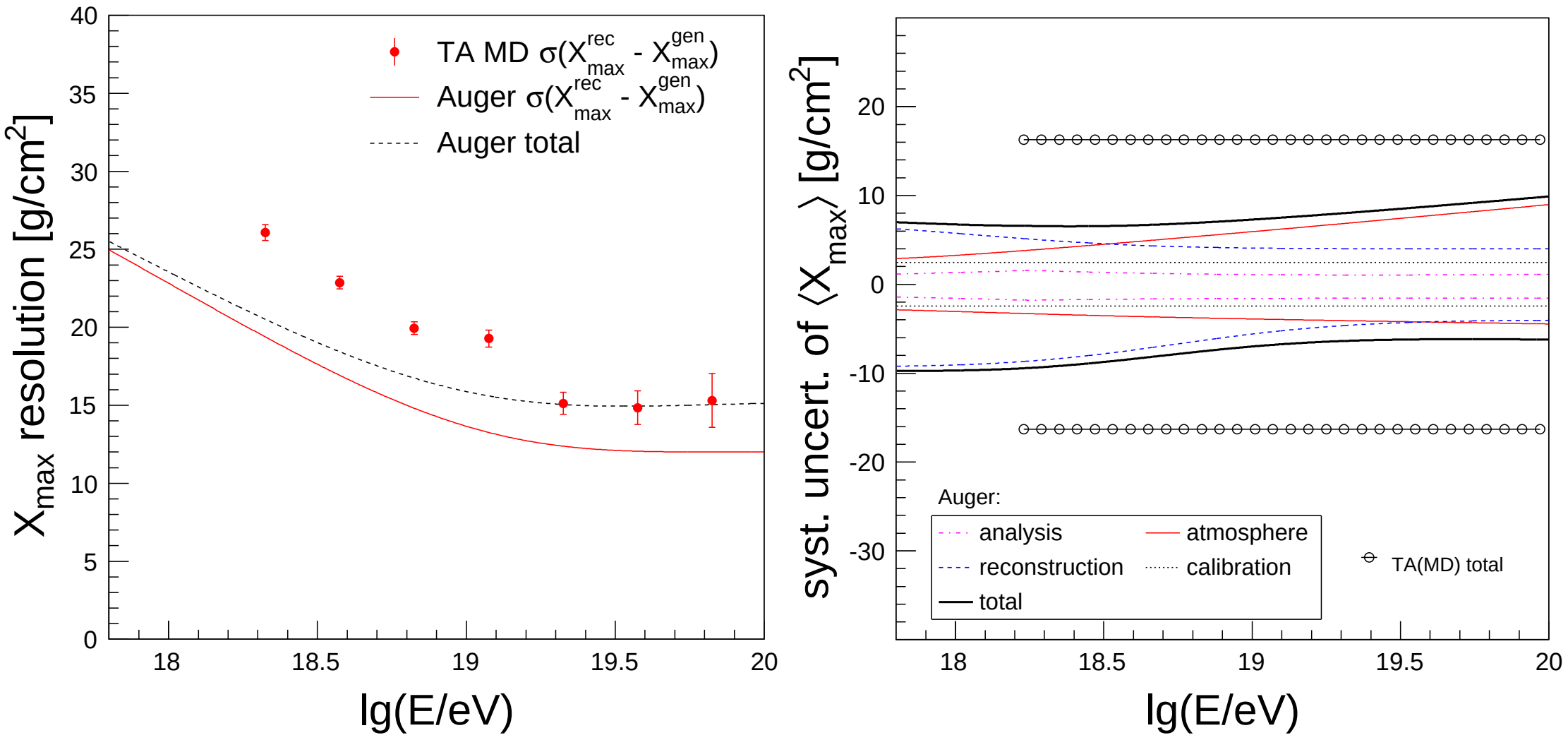

**Figure 3:** $X_{\max}$ resolution (left) and systematics of the $X_{\max}$ scale (right) for the Auger and TA analyses.

Technically, the parametric description of the $X_{\max}$ distribution is realized by providing a set of composition fractions as a function of energy that describe the $X_{\max}$ distributions measured by Auger. These fractions are obtained as described in [5] by a log-likelihood fit of templates of $X_{\max}$ distributions for different nuclear primaries as predicted by air-shower simulations using a particular hadronic interaction model. It is worthwhile noting that the detector acceptance and resolution at a given primary energy depend mainly on $X_{\max}$ itself and only weakly on the primary particle type or hadronic interaction model via the invisible energy. Therefore, for the analysis presented here, it is only important that the resulting composition mix describes the data well and not which fractions of primaries are needed or which hadronic interaction model is used to obtain a model of the undistorted $X_{max}$ distribution based on Auger data.

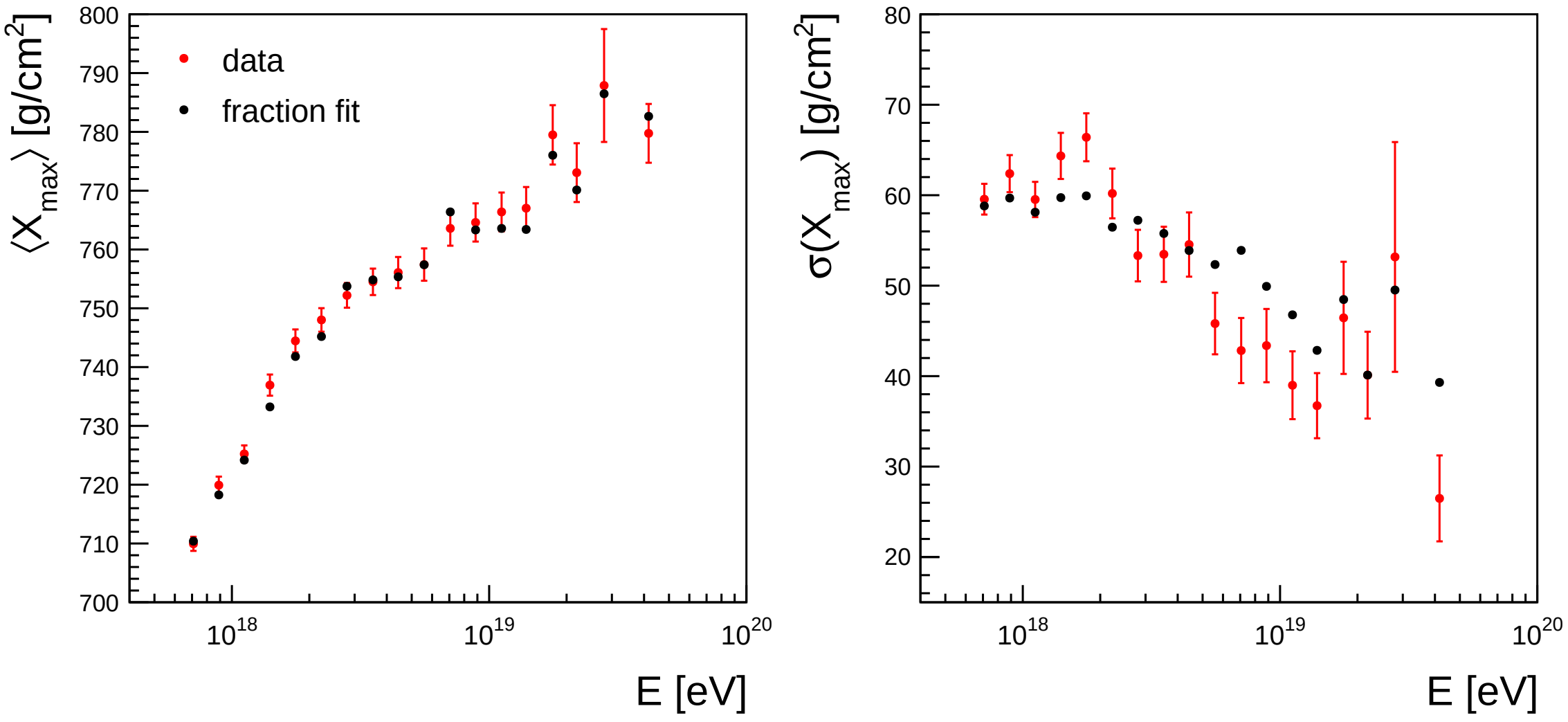

**Figure 4:** Moments of the fitted $X_{max}$ distributions using QGSJETII-03 (black markers) and $X_{max}$ moments measured by the Pierre Auger Collaboration (red circles with statistical error bars), see text.

Here we used QGSJETII-03 [6] and a mix of four primary particles (proton, helium, nitrogen and iron) to obtain a model of the true $X_{max}$ distribution based on the Auger data. QGSJETII-03 is not included in the set of models studied by the Pierre Auger Collaboration to infer the primary composition [5] because it gives a worse description of LHC data than the re-tuned version QGSJETII-04 [7]. However, with neither version of QGSJETII it is possible to find a composition mix that gives a perfect description of the $X_{max}$ distributions measured by Auger. The first two moments of the best fits with QGSJETII-03 and the Auger data are shown in Fig. 4. As can be seen, there is a good agreement regarding $\langle X_{max} \rangle$, but there are deviations between the fitted and observed width of the distribution.

Ideally, this analysis should be performed with a combination of composition and hadronic interaction model that fits the Auger data well, such as SIBYLL2.1 [8] or EPOS-LHC [9] (see discussion in [5]). However, for practical reasons, we performed a preliminary analysis with QGSJETII-03. Since the deviations between the moments of the data and the ones of the fitted distributions are on average at the $5\,\mathrm{g/cm^2}$ level, this approach is expected to give only a small bias in the predictions for the observed distributions.

In detail, the analysis proceeds as follows: the composition mix is processed using the hybrid-reconstruction-analysis software of the Telescope Array. Showers are generated with CORSIKA and the trigger response of the scintillator array is simulated. The longitudinal shower profile from CORSIKA is fitted to a Gaisser-Hillas function to determine the shower parameters and the fitted profile is used consecutively to generate the light emission. The TA fluorescence detector response including atmospheric, electronics, and geometrical acceptance is then simulated. Subsequently the event geometry is fitted via the fluorescence profile, and the shower-detector plane is measured. A fit to hybrid shower geometry is performed which combines the timing and geometric center of charge of the surface detector array, with the timing and geometry of the fluorescence detector that observed the event. This step is what makes the event a "hybrid event". If either the surface or fluorescence detector fail to trigger in an event, it is not processed any further, otherwise the shower profile is fitted via a reverse Monte Carlo method where the atmosphere, electronics, and geometrical acceptance of the shower are fully simulated.

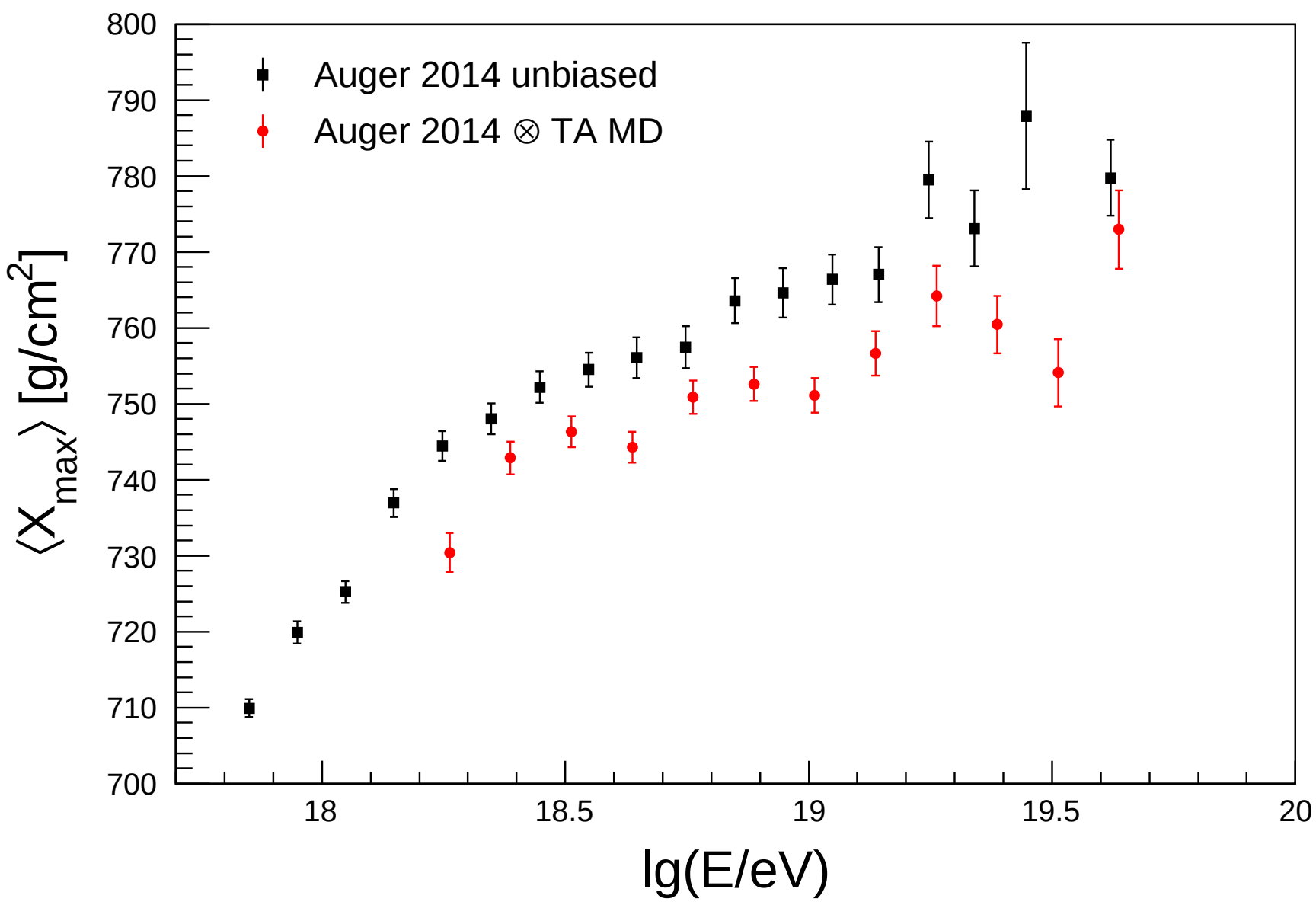

**Figure 5:** Effect of the MD detector acceptance on $X_{\max}$. The $\langle X_{\max}\rangle$ of an $X_{\max}$ distribution describing the Auger data before and after the MD acceptance are shown as solid squares and circles respectively. The error bars denote the statistical uncertainties of the Auger result in case of the squares and the statistical uncertainties due to the limited MC statistics in the case of the circles.

The resulting effect of the folding of the parametric Auger distributions with the TA detector response, reconstruction and analysis on the $\langle X_{\max}\rangle$ of Auger is shown in Fig. 5. As can be seen, the mean value after the application of the TA detector response is smaller than the generated mean.

## 4. Results and Discussion

The $\langle X_{\max}\rangle$ as measured by TA using the MD fluorescence telescope and the Auger result folded with the TA acceptance are shown in Fig. 6. Their compatibility is quantified with a bin-by-bin comparison excluding the highest-energy data points of each experiment which are at different energies. Using only the statistical uncertainties yields a $\chi^2/\mathrm{Ndf}$ of 10.7/11 with $P(\chi^2 \geq 10.7|11) = 0.47$. The average difference of the data points is $(2.9 \pm 2.7\ (\mathrm{stat.}) \pm 18\ (\mathrm{syst.}))\ \mathrm{g/cm}^2$ with a $\chi^2/\mathrm{Ndf}$ of 9.5/10 ($P = 0.48$). It can be concluded that the two data sets are in excellent agreement, even without accounting for the respective systematic uncertainties on the $X_{\max}$ scale. However, in the present study we did not take into account a possible difference in the energy scale of the two experiments. The comparison of the energy spectra at the ankle region suggests that the energy scale of TA is about 13% higher than the one of the Pierre Auger Observatory [10].

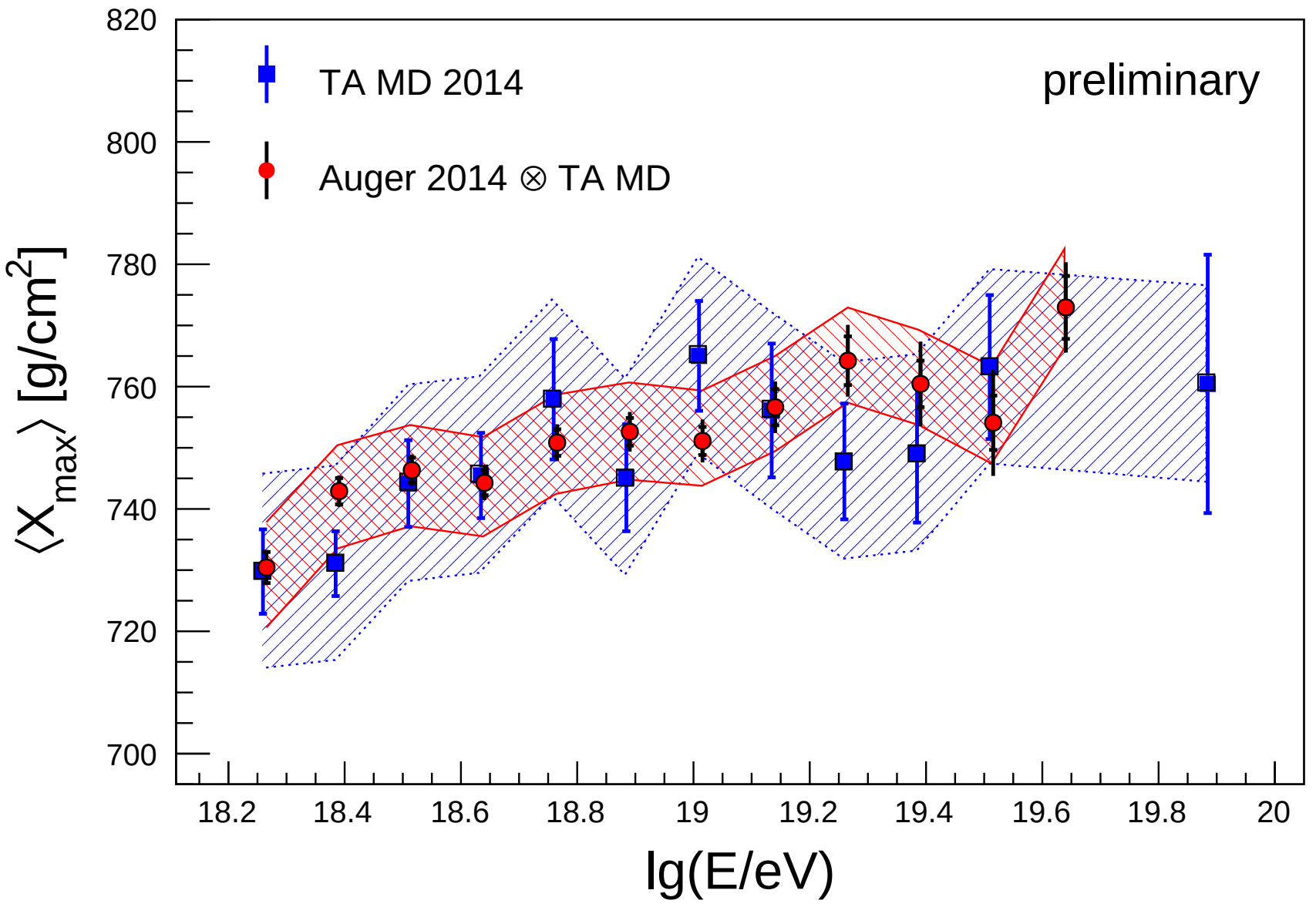

**Figure 6:** Comparison of $\langle X_{\rm max}\rangle$ as measured with the MD of TA (blue squares) and the $\langle X_{\rm max}\rangle$ of the Auger data folded with the MD acceptance (red circles). The data points are slightly shifted horizontally for better visibility. In the case of the Auger points, the inner error bars denote the statistical uncertainty of the measurement and the total error bar also includes contributions from the limited statistics of simulated events used for the folding. The colored bands show the systematic uncertainties of the $X_{\rm max}$ scales of each experiment.

However, since the elongation rate of the folded Auger data is small ($\sim$19 g/cm$^2$/decade), the effect of such an energy shift on the comparison is expected to be at the level of a few g/cm$^2$. For a more precise evaluation it would be necessary to take into account the energy dependence of the acceptance of TA. Nevertheless, it is to be expected that the increased difference between the two data sets once the energy scale shift is taken into account will be much smaller than the systematic uncertainties on the $X_{\rm max}$ scale of $\leq$10 g/cm$^2$ and 16 g/cm$^2$ for the Auger and TA analyses respectively.

## 5. Conclusions and Outlook

In this paper we have presented a comparison between the data on $\langle X_{\rm max}\rangle$ as measured by the Pierre Auger and Telescope Array Collaborations. An adequate comparison was achieved by taking into account that the $\langle X_{\rm max}\rangle$ published by Auger are corrected for detector effects, whereas those published by TA includes detector effects. From the preliminary comparison presented here we conclude that the data of the two observatories are in good agreement.

In the future, we will present results with an improved parametric description of the Auger $X_{\rm max}$ distributions using the EPOS-LHC interaction model and the evaluation of the effect of the relative energy scale uncertainty. Moreover, we will discuss results from statistical tests of the compatibility of the full $X_{\rm max}$ distribution.

## References

[1] E. Barcikowski *et al.* for the Pierre Auger, Telescope Array and Yakutsk Collaborations, *Mass Composition Working Group Report at UHECR-2012*, *EPJ Web Conf.* **53** (2013) 01006

[2] A. Aab *et al.* [Pierre Auger Collaboration], *Depth of maximum of air-shower profiles at the Pierre Auger Observatory. I. Measurements at energies above $10^{17.8}$ eV*, *Phys. Rev. D* **90** (2014) 122005

[3] R.U. Abbasi *et al.* [Telescope Array Collaboration], *Study of Ultra-High Energy Cosmic Ray composition using Telescope Array's Middle Drum detector and surface array in hybrid mode*, *Astropart. Phys.* **64** (2014) 49.

[4] W.F. Hanlon for the Pierre Auger and Telescope Array Collaborations, *Progress towards understanding the analyses of mass composition made by the Auger and Telescope Array Collaborations*, *Proc. 33rd ICRC* (2013) [`arXiv:1310.0647`].

[5] A. Aab *et al.* [Pierre Auger Collaboration], *Depth of maximum of air-shower profiles at the Pierre Auger Observatory. II. Composition implications*, *Phys. Rev. D* **90** (2014) 122006.

[6] S. Ostapchenko, *Monte Carlo treatment of hadronic interactions in enhanced Pomeron scheme: I. QGSJET-II model*, *Phys. Rev. D* **83** (2011) 014018.

[7] S. Ostapchenko, *QGSJET-II: physics, recent improvements, and results for air showers*, *EPJ Web Conf.* **52** (2013) 02001.

[8] E.J. Ahn, R. Engel, T.K. Gaisser, P. Lipari and T. Stanev, *Cosmic ray interaction event generator SIBYLL 2.1*, *Phys. Rev. D* **80** (2009) 094003.

[9] T. Pierog, I. Karpenko, J.M. Katzy, E. Yatsenko and K. Werner, *EPOS LHC : test of collective hadronization with LHC data* (2013) [`arXiv:1306.0121`].

[10] I.C. Mariş for the Pierre Auger and Telescope Array Collaborations, *High Energy Spectrum Working Group Report*, presentation at the UHECR symposium (2014), Springdale, USA, http://uhecr2014.telescopearray.org/maris/TAAuger_Springdale.pdf

# Large-Scale Distribution of Arrival Directions of Cosmic Rays Detected at the Pierre Auger Observatory and the Telescope Array above $10^{19}$ eV

**Olivier Deligny**[*a] **for the Pierre Auger and Telescope Array Collaborations**[b,c]
[a] *CNRS/IN2P3 - IPN Orsay, France*
[b] *Observatorio Pierre Auger, Av. San Martín Norte 304, 5613 Malargüe, Argentina*
*E-mail:* auger_spokespersons@fnal.gov
*Full author list:* http://www.auger.org/archive/authors_2015_ICRC.html
[c] *Telescope Array Project, 201 James Fletcher Bldg, 115 S. 1400 East, Salt Lake City, UT 84112-0830, USA*
*E-mail:* ta-icrc@cosmic.utah.edu
*Full author list:*
http://www.telescopearray.org/index.php/research/publications/conference-proceedings

The large-scale distribution of arrival directions of high-energy cosmic rays is a key observable in attempts to understanding their origin. The dipole and quadrupole moments are of special interest in revealing potential anisotropies. An unambiguous measurement of these moments as well as of the full set of spherical harmonic coefficients requires full-sky coverage. This can be achieved by combining data from observatories located in both the northern and southern hemispheres. To this end, a joint analysis using data recorded at the Pierre Auger Observatory and the Telescope Array above $10^{19}$ eV has been performed. Thanks to the full-sky coverage, the measurement of the dipole moment reported in this study does not rely on any assumption on the underlying flux of cosmic rays. As well, the resolution on the quadrupole and higher order moments is the best ever obtained. The resulting multipolar expansion of the flux of cosmic rays allows a series of anisotropy searches to be performed, and in particular to report on the first angular power spectrum of cosmic rays. This allows a comprehensive description of the angular distribution of cosmic rays above $10^{19}$ eV.

*Speaker.

*Introduction.* Above $10^{19}$ eV, the flux of Ultra-High Energy Cosmic Rays (UHECRs) is expected to be of extragalactic origin. Although the actual sources of UHECRs are still to be identified, their distribution in the sky is expected to follow, to some extent, the large-scale structure of the matter in the Universe. Due to scattering in magnetic fields, the anisotropy imprinted upon the distribution of arrival directions is mainly expected at large scales even for energies as large as $10^{19}$ eV. A non-zero dipole moment is naturally expected for diffusive propagation of UHECRs from nearby sources, leading to a cosmic-ray density gradient embedding the observer. On the other hand, excesses along a plane, for instance the super-Galactic one, would be detectable as a prominent quadrupole. The dipole and the quadrupole moments are thus of special interest, but an access to the full set of multipoles is relevant to characterize departures from isotropy at all scales.

Recently, the full-sky multipole coefficients of the UHECR flux have been measured for the first time by the Pierre Auger and Telescope Array collaborations using a joint data set with energies above $10^{19}$ eV [1]. No significant deviations from isotropic expectations were found, and upper limits on the amplitudes of the dipole and quadrupole moments were reported as a function of the direction in the sky, varying between 8% and 13% for the dipole and between 7.5% and 10% for a symmetric quadrupole. In the meanwhile, another recent report from the Pierre Auger Collaboration based on an enlarged data set has indicated an amplitude for the first harmonic in right ascension of $(4.4 \pm 1.0)\%$ with a chance probability of $6.4 \times 10^{-5}$ for events with energies larger than $8 \times 10^{18}$ eV [2]. Under the assumption that the only significant contribution to the anisotropy is from the dipolar component, this observation corresponds to a dipole of amplitude $(7.3 \pm 1.5)\%$ pointing to $(95^\circ \pm 13^\circ)$ in right ascension and $(-39^\circ \pm 13^\circ)$ in declination.

The aim of this joint study is to search for anisotropy with full-sky coverage in a similar way as reported in [1], by including one additional year of data recorded at the Telescope Array and by extending the zenithal range of the data recorded at the Pierre Auger Observatory from $[0^\circ - 60^\circ]$ to $[0^\circ - 80^\circ]$. A detailed description of the estimation of spherical harmonic coefficients when combining data from Auger and Telescope Array is available in [1], so that only basic principles are summarized in this report. The extension of the zenithal range in the Auger data set provides a larger overlap of the field of view of the two experiments compared to the previous report. This allows a significant improvement of the resolution on the dipole moment. The joint data set considered in this study consists of events above $10^{19}$ eV recorded at the Telescope Array from May 2008 to May 2014 with an exposure of $7,250$ km$^2$ yr sr with zenith angles less than $55^\circ$, and at the Pierre Auger Observatory from 1 Jan. 2004 to 31 Dec. 2013 with an exposure of $48,029$ km$^2$ yr sr with zenith angles less than $80^\circ$.

*The full-sky directional exposure.* The directional exposure $\omega(\mathbf{n})$ provides the effective time-integrated collecting area for a flux from each direction of the sky. Since the energy threshold of $10^{19}$ eV guarantees that both experiments are fully efficient in their respective zenith range, the directional exposure relies only on the geometrical acceptance [3]. The functions $\omega_i(\delta)$ of each experiment are shown in fig. 1. Given the respective latitudes of both observatories and with the maximum zenith angle used here, full-sky coverage is naturally achieved when summing both functions. Also, it is to be noted that a common band of declination, namely $-15^\circ \leq \delta \leq 45^\circ$, is covered by both experiments.

In principle, the combined directional exposure of the two experiments should be simply the sum of the individual ones. However, individual exposures have here to be re-weighted by some factor $b$ due to the unavoidable uncertainty in the relative exposures of the experiments:

$$\omega(\mathbf{n};b) = \omega_{\mathrm{TA}}(\mathbf{n}) + b\,\omega_{\mathrm{Auger}}(\mathbf{n}). \tag{1}$$

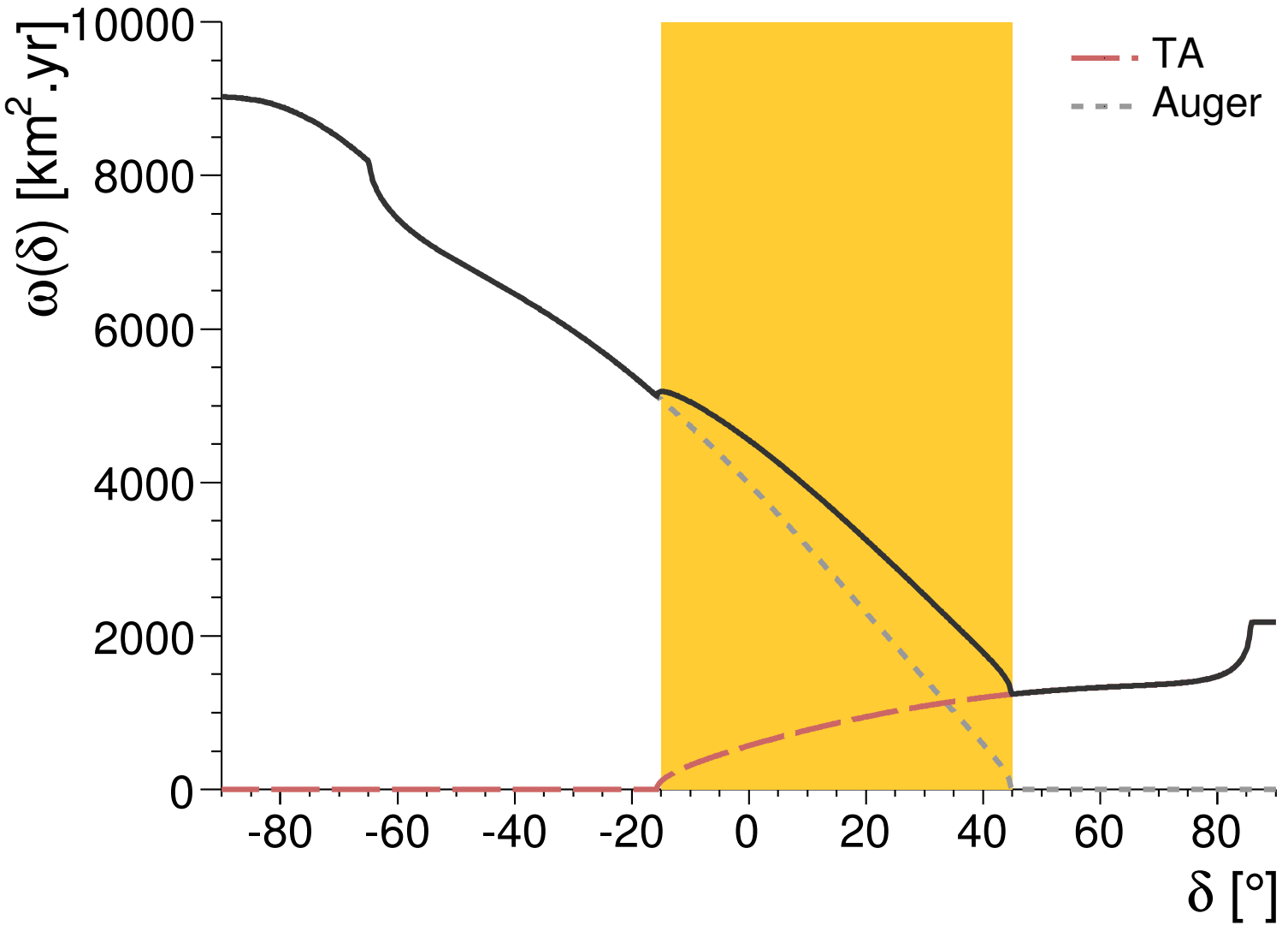

**Figure 1:** Directional exposure above $10^{19}$ eV as obtained by summing the nominal individual ones of the Telescope Array and the Pierre Auger Observatory, as a function of the declination. The overlapping sky region is indicated by the yellow band.

Written in this way, $b$ is a dimensionless parameter of order unity arbitrarily chosen to re-weight the directional exposure of the Pierre Auger Observatory relative to the one of the Telescope Array. In practice, only an estimation $\bar{b}$ of the factor $b$ can be obtained, so that only an estimation of the directional exposure $\bar{\omega}(\mathbf{n}) \equiv \omega(\mathbf{n};\bar{b})$ can be achieved through eqn. 1. In addition, although the techniques for assigning energies to events are nearly the same, there are differences as to how the primary energies are derived at the Pierre Auger Observatory and the Telescope Array. Currently, systematic uncertainties in the energy scale of both experiments amount to about 14% and 21% respectively [4, 5]. Such a potential shift in energy leads to different counting rates above some fixed energy threshold, which induces fake anisotropies in a similar way to the ones resulting from a shift in the relative exposures of the experiments. The parameter $b$ can thus be viewed as an effective correction which absorbs any kind of systematic uncertainties in the relative exposures, whatever the sources of these uncertainties [1].

*Estimation of the spherical harmonic coefficients.* The flux of cosmic rays $\Phi(\mathbf{n})$ can be decomposed in terms of a multipolar expansion onto the spherical harmonics $Y_{\ell m}(\mathbf{n})$,

$$\Phi(\mathbf{n}) = \sum_{\ell\geq 0}\sum_{m=-\ell}^{\ell} a_{\ell m} Y_{\ell m}(\mathbf{n}). \tag{2}$$

[1]Note that variations of the exposure with time due to unavoidable changes in the experimental conditions induce a dependence of the exposure functions in right ascension. These variations are however neglected in this analysis, given the much larger statistical uncertainties due to the overall small number of events above $10^{19}$ eV.

Any anisotropy fingerprint is encoded in the multipoles $a_{\ell m}$. Non-zero amplitudes in the $\ell$ modes arise from variations of the flux on an angular scale $\simeq 1/\ell$ radians. The observed angular distribution of cosmic rays, $\mathrm{d}N/\mathrm{d}\Omega$, can be naturally modeled as the sum of Dirac functions on the surface of the unit sphere the arguments of which are the arrival directions $\{\mathbf{n}_1, \ldots, \mathbf{n}_N\}$ of the events,

$$\frac{\mathrm{d}N(\mathbf{n})}{\mathrm{d}\Omega} = \sum_{i=1}^{N} \delta(\mathbf{n}, \mathbf{n}_i). \tag{3}$$

Here, arrival directions are expressed in the equatorial coordinate system (declination $\delta$ and right ascension $\alpha$) since this is the most natural one tied to the Earth in describing the directional exposure of any experiment. With full-sky but non-uniform coverage, the customary recipe for decoupling directional exposure effects from anisotropy ones consists in weighting the observed angular distribution by the inverse of the *relative* directional exposure function [3]:

$$\frac{\mathrm{d}\tilde{N}(\mathbf{n})}{\mathrm{d}\Omega} = \frac{1}{\bar{\omega}_r(\mathbf{n})} \frac{\mathrm{d}N(\mathbf{n})}{\mathrm{d}\Omega}. \tag{4}$$

The relative directional exposure $\bar{\omega}_r$ is a dimensionless function normalized to unity at its maximum. In turn, when combining the exposure of the two experiments with an unbiased estimator of $b$, the recovered coefficients defined as

$$\bar{a}_{\ell m} = \int_{4\pi} \mathrm{d}\Omega \frac{\mathrm{d}\tilde{N}(\mathbf{n})}{\mathrm{d}\Omega} Y_{\ell m}(\mathbf{n}) = \sum_{i=1}^{N} \frac{Y_{\ell m}(\mathbf{n}_i)}{\bar{\omega}_r(\mathbf{n}_i)}, \tag{5}$$

provide unbiased estimators of the underlying $a_{\ell m}$ multipoles. Under reasonable assumptions, the resolution $\sigma_{\ell m}$ on each $a_{\ell m}$ multipole can be shown to be the sum of a first term reflecting the Poisson fluctuations induced by the finite number of events and of a second term reflecting the uncertainty in the relative exposures of the two experiments through the uncertainty to estimate the $b$ parameter. This term mainly impacts the resolution in the dipole coefficient $a_{10}$, while it has a small influence on the quadrupole coefficient $a_{20}$ and a marginal one on higher order moments $\{a_{\ell 0}\}_{\ell \geq 3}$.

*The Joint-Analysis Method.* The band of declinations between $-15^\circ$ and $45^\circ$ is exposed to the fields of view of both experiments. This overlapping region can be used for designing a procedure to get simultaneously a relevant estimate of the parameter $b$ and of the multipole coefficients $a_{\ell m}$ through an iteration method.

Considering as a first approximation the flux $\Phi(\mathbf{n})$ as isotropic, the overlapping region denoted by $\Delta\Omega$ can be utilized to derive a first estimate $\bar{b}^{(0)}$ of the $b$ factor by requiring the observed fluxes of both experiments to be identical in this particular region:

$$\bar{b}^{(0)} = \frac{\Delta N_{\mathrm{Auger}}}{\Delta N_{\mathrm{TA}}} \frac{\int_{\Delta\Omega} \mathrm{d}\Omega\, \omega_{\mathrm{TA}}(\mathbf{n})}{\int_{\Delta\Omega} \mathrm{d}\Omega\, \omega_{\mathrm{Auger}}(\mathbf{n})}, \tag{6}$$

with $\Delta N_{\mathrm{Auger}}$ and $\Delta N_{\mathrm{TA}}$ the number of events observed in the overlapping region in each experiment. Then, inserting $\bar{b}^{(0)}$ into $\bar{\omega}$, 'zero-order' $\bar{a}_{\ell m}^{(0)}$ coefficients can be obtained. This set of coefficients is only a rough estimation, due to the limiting assumption on the flux (isotropy).

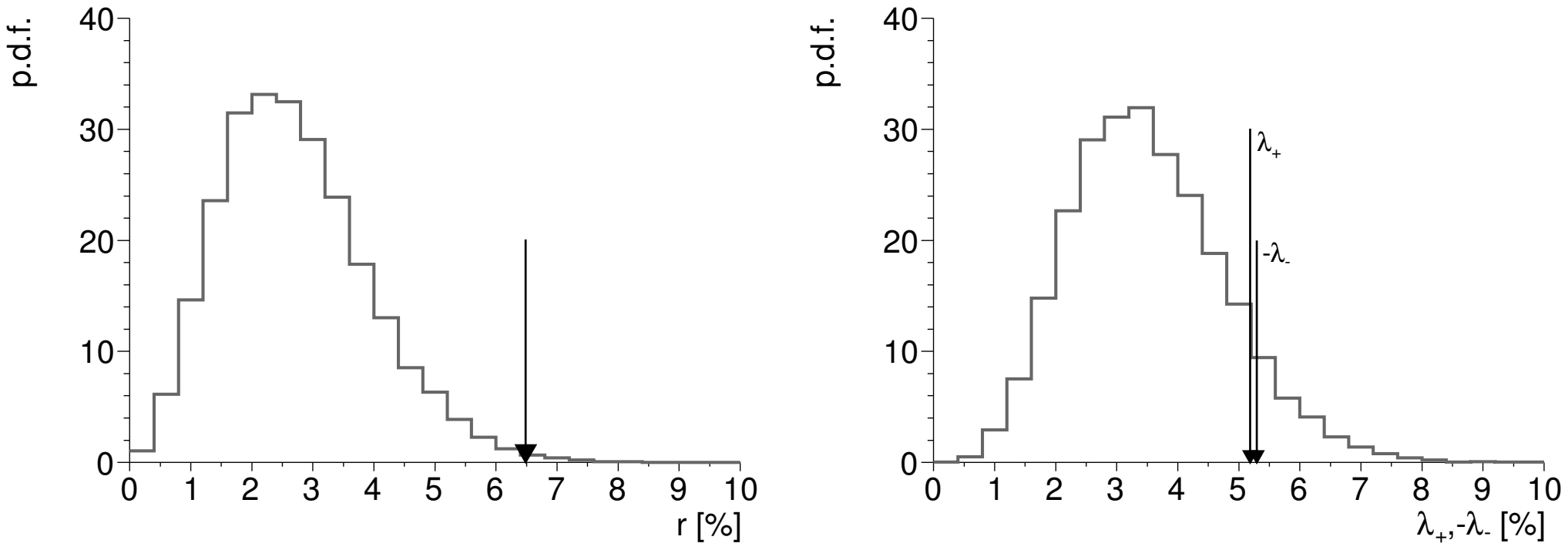

**Figure 2:** Measured amplitudes for the dipole vector (left) and the quadrupole tensor (right), together with the distributions expected from statistical fluctuations of isotropy.

On the other hand, the expected number of events in the common band for each observatory, $\Delta N_{\mathrm{TA}}^{\mathrm{exp}}$ and $\Delta N_{\mathrm{Auger}}^{\mathrm{exp}}$, can be expressed from the underlying flux $\Phi(\mathbf{n})$ as

$$\begin{aligned}\Delta N_{\mathrm{TA}}^{\mathrm{exp}} &= \int_{\Delta\Omega} \mathrm{d}\Omega\, \Phi(\mathbf{n})\, \omega_{\mathrm{TA}}(\mathbf{n}), \\ \Delta N_{\mathrm{Auger}}^{\mathrm{exp}} &= \int_{\Delta\Omega} \mathrm{d}\Omega\, \Phi(\mathbf{n})\, \omega_{\mathrm{Auger}}(\mathbf{n}).\end{aligned} \tag{7}$$

From eqns. 7, and from the set of $\bar{a}_{\ell m}^{(0)}$ coefficients, an iterative procedure estimating at the same time $b$ and the set of $a_{\ell m}$ coefficients can be constructed as

$$\bar{b}^{(k+1)} = \frac{\Delta N_{\mathrm{Auger}}}{\Delta N_{\mathrm{TA}}} \frac{\int_{\Delta\Omega} \mathrm{d}\Omega\, \bar{\Phi}^{(k)}(\mathbf{n})\, \omega_{\mathrm{TA}}(\mathbf{n})}{\int_{\Delta\Omega} \mathrm{d}\Omega\, \bar{\Phi}^{(k)}(\mathbf{n})\, \omega_{\mathrm{Auger}}(\mathbf{n})}, \tag{8}$$

where $\bar{\Phi}^{(k)}$ is the flux estimated with the set of $\bar{a}_{\ell m}^{(k)}$ coefficients. This iterative procedure has been shown to provide unbiased estimators of the multipole coefficients without any assumptions on the flux [1].

Note that the resolution on the $b$ parameter turns out to be 2.1% in the present study, compared to 3.9% in the previous report [1]. This offers a much better resolution in estimating the $a_{10}$ coefficient. Part of this improvement is due to the statistics increased by $\simeq 46\%$. The other factor is provided by the extended declination band exposed to the Auger and Telescope Array fields of view thanks to the use of horizontal events in the Auger data set.

*Low-order multipoles.* All analyses reported hereafter are based on a joint data set consisting of events with energies in excess of $10^{19}$ eV in terms of the energy scale used at the Telescope Array by evaluating in the Auger data set the energy threshold which guarantees equal fluxes for both experiments. We are thus left here with 2,560 events (1,703 in the common band) above $10^{19}$ eV from the Telescope Array and 16,835 (5,885 in the common band) above $8.8 \times 10^{18}$ eV from the Pierre Auger Observatory. After iterations, the coefficient $b$ is $b = 0.96$.

Although the full set of spherical harmonic moments is needed to characterise any departure from isotropy at any scale, the dipole and quadrupole moments are of special interest. For that reason, a special emphasis is given here to these low-order moments, in terms of a more traditional and geometric representation than the raw result of the multipole moments. The dipole moment can be fully characterized by a vector with an amplitude $r$ and the two angles $\{\delta_\mathrm{d}, \alpha_\mathrm{d}\}$ of the unit vector $\mathbf{d}$. The quadrupole, on the other hand, can be fully determined by two independent amplitudes $\{\lambda_+, \lambda_-\}$, two angles $\{\delta_{\mathrm{q}_+}, \alpha_{\mathrm{q}_+}\}$ defining the orientation of a unit vector $\mathbf{q}_+$, and one additional angle $\alpha_{\mathrm{q}_-}$ defining the directions of another unit vector $\mathbf{q}_-$ in the orthogonal plane to $\mathbf{q}_+$. The full description is completed by means of a third unit vector $\mathbf{q}_0$, orthogonal to both $\mathbf{q}_+$ and $\mathbf{q}_-$, and with a corresponding amplitude such that the traceless condition $\lambda_+ + \lambda_0 + \lambda_- = 0$ is satisfied. The parameterisation of the low-order moments of the flux is then written in a convenient and intuitive way as

$$\Phi(\mathbf{n}) = \frac{\Phi_0}{4\pi}\left(1 + r\,\mathbf{d}\cdot\mathbf{n} + \lambda_+(\mathbf{q}_+\cdot\mathbf{n})^2 + \lambda_0(\mathbf{q}_0\cdot\mathbf{n})^2 + \lambda_-(\mathbf{q}_-\cdot\mathbf{n})^2 + \cdots\right). \tag{9}$$

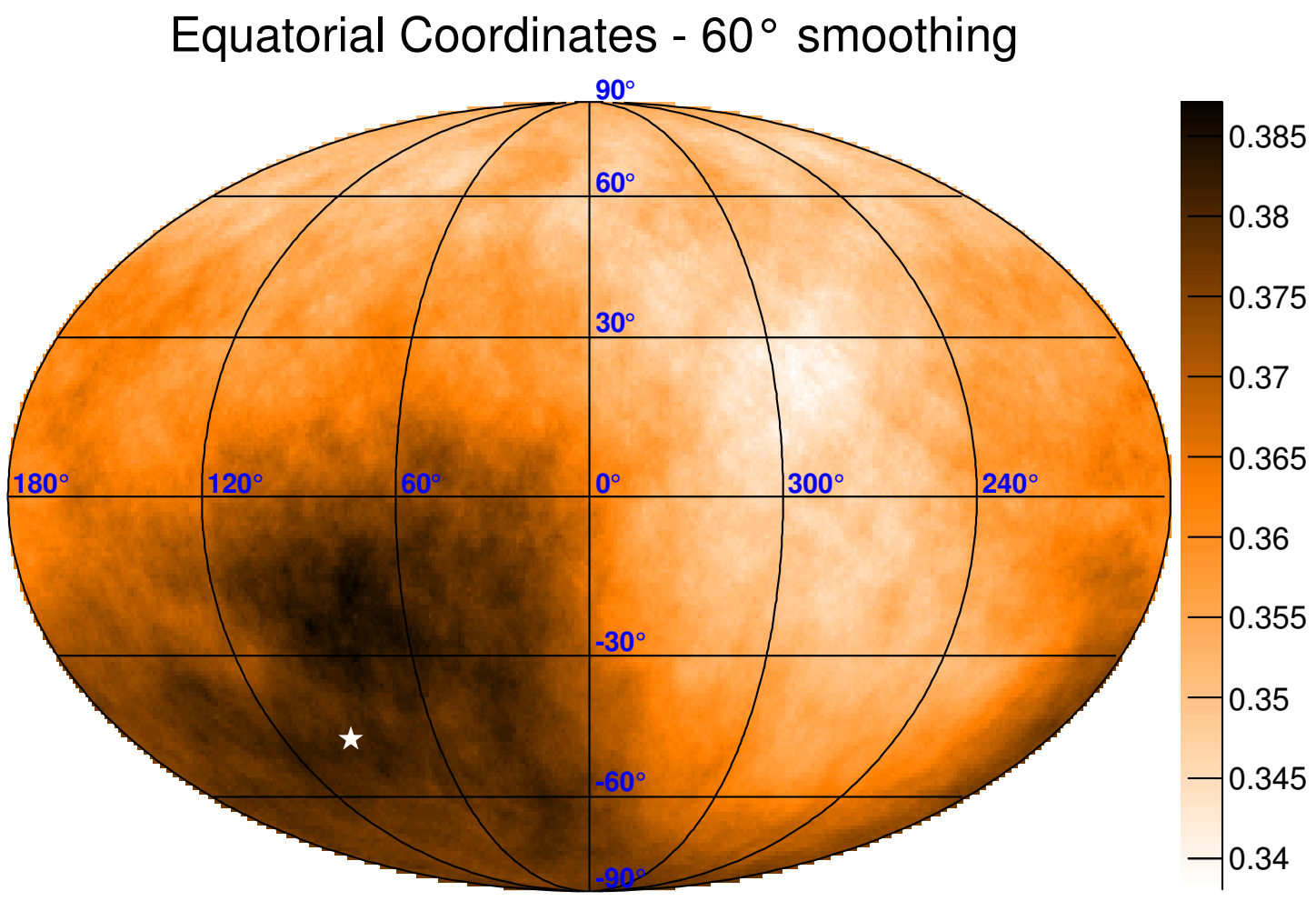

**Figure 3:** Sky map in equatorial coordinates of the average flux smoothed out at a 60° angular scale above $10^{19}$ eV in km$^{-2}$ yr$^{-1}$ sr$^{-1}$ units.

The distributions of amplitudes obtained from statistical fluctuations of simulated isotropic samples are shown in fig. 2. The measured values are indicated by the superimposed arrows. The dipole amplitude is observed to be $(6.5 \pm 1.9)\%$ with a chance probability of $5 \times 10^{-3}$, pointing to $(93^\circ \pm 24)$ in right ascension and $(-46^\circ \pm 18)$ in declination. Compared to the previous report in [1], the improved sensitivity in the dipole moment is primarily explained by the improved resolution on the $b$ parameter thanks to the larger common band $\Delta\Omega$, and by the increased exposure/statistics. On the other hand, the quadrupole amplitudes are observed to be within statistical fluctuations expected from isotropic samples. Overall, these results are in agreement with the ones reported in [2] without any assumption on the underlying flux of UHECRs.

To visualise the recovered dipole moment, an average flux smoothed out at an angular scale $\Theta$ per solid angle unit can be derived using the joint data set in the following way:

$$\langle\Phi(\mathbf{n})\rangle_\Theta = \frac{1}{\int_\Theta \mathrm{d}\mathbf{n}} \int_\Theta \mathrm{d}\mathbf{n}'\, f(\mathbf{n},\mathbf{n}')\, \frac{1}{\bar{\omega}(\mathbf{n}')}\, \frac{\mathrm{d}N(\mathbf{n}')}{\mathrm{d}\Omega}, \tag{10}$$

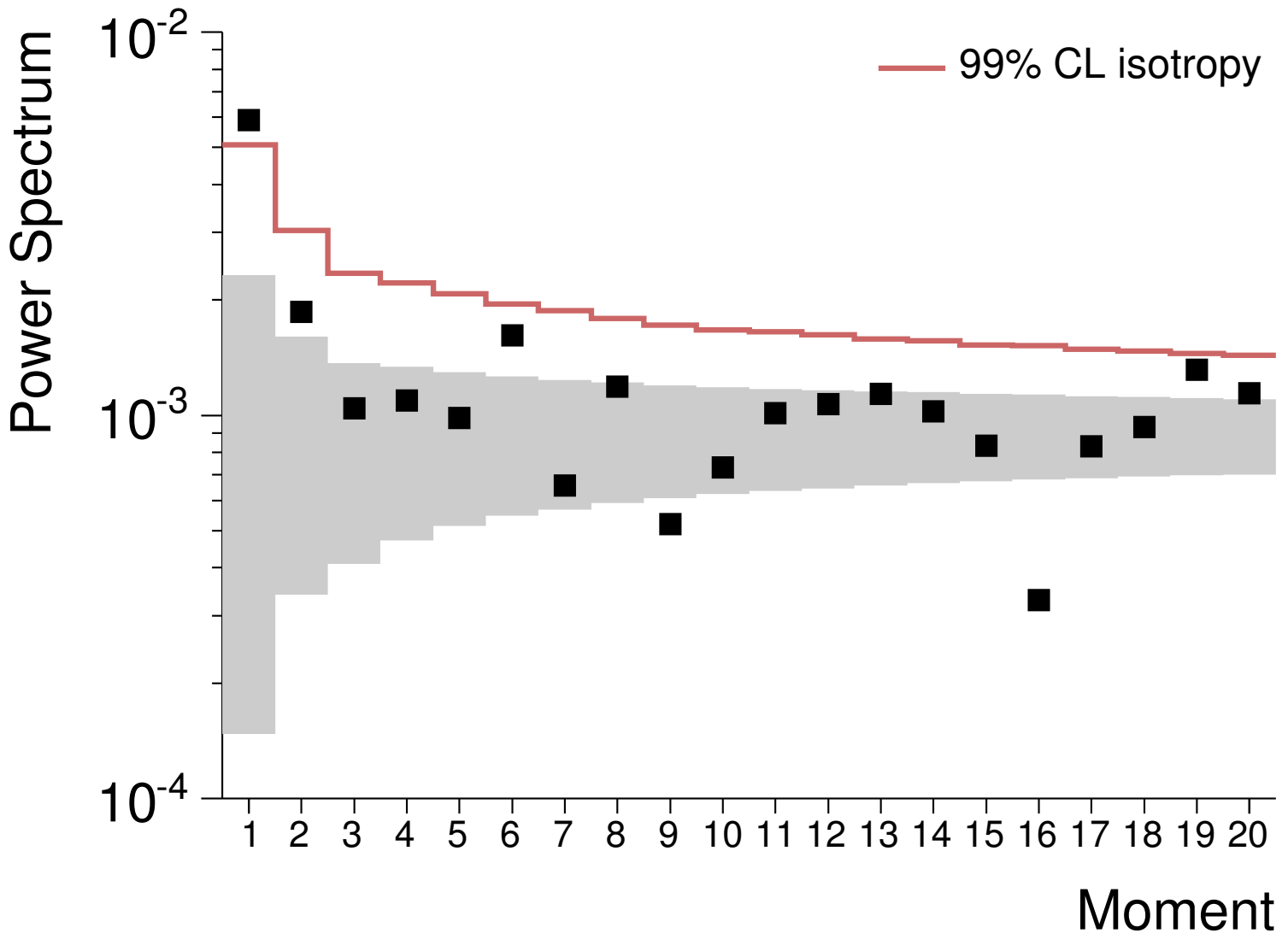

**Figure 4:** Angular power spectrum.

with $f$ the top-hat filter function at the angular scale $\Theta$. This average flux is displayed using the Mollweide projection in fig. 3, in km$^{-2}$yr$^{-1}$sr$^{-1}$ units. This map is drawn in equatorial coordinates. To exhibit the dipole structure, the angular window is chosen to be $\Theta = 60^\circ$. The direction of the reconstructed dipole is shown as the white star.

*Angular power spectrum.* The angular power spectrum $C_\ell$ is a coordinate-independent quantity, defined as the average $|a_{\ell m}|^2$ as a function of $\ell$,

$$C_\ell = \frac{1}{2\ell+1}\sum_{m=-\ell}^{\ell} |a_{\ell m}|^2 . \tag{11}$$

In the same way as the multipole coefficients, any significant anisotropy of the angular distribution over scales near $1/\ell$ radians would be captured in a non-zero power in the mode $\ell$. Although the exhaustive information of the distribution of arrival directions is encoded in the full set of multipole coefficients, the characterisation of any important overall property of the anisotropy is hard to handle in a summary plot from this set of coefficients. Conversely, the angular power spectrum does provide such a summary plot. In addition, it is possible that for some fixed mode numbers $\ell$, all individual $a_{\ell m}$ coefficients do not stand above the background noise but meanwhile do so once summed quadratically.

From the set of estimated coefficients $\bar{a}_{\ell m}$, the measured power spectrum is shown in fig. 4. The gray band stands for the RMS of power around the mean values expected from an isotropic distribution, while the solid line stands for the 99% confidence level upper bounds that would result from fluctuations of an isotropic distribution. The dipole moment is observed to stand out from the background noise, with a chance probability of $5\times10^{-3}$. Beyond the dipole, no other multipole deviates from expected fluctuations at 99% CL in an isotropic flux.

*Outlook.* In this work, an entire mapping of the celestial sphere has been achieved by combining data sets recorded at the Pierre Auger Observatory and the Telescope Array above $10^{19}$ eV. Throughout the series of anisotropy searches performed, a dipole moment with an amplitude $r = (6.5 \pm 1.9)\%$ is captured with a chance probability of $5 \times 10^{-3}$, while no other deviation from isotropy can be revealed at smaller angular scales.

Large-scale anisotropies of cosmic rays with energies in excess of $10^{19}$ eV are closely connected to the sources and the propagation mode of extragalactic UHECRs, see e.g. [8, 9]. Due to scattering in the extragalactic magnetic fields, large deflections are expected even at such high energies for field amplitudes ranging in few nanogauss and extended over coherence lengths of the order of one megaparsec, or even for lower amplitudes if the electric charge of UHECRs is large. For sources distributed in a similar way to the matter in the Universe, the angular distribution of UHECRs is then expected to be influenced by the contribution of nearby sources, so that the Milky Way should be embedded into a density gradient of cosmic rays that should lead to at least a dipole moment. The contribution of nearby sources is even expected to become dominant as the energy of cosmic rays increases due to the reduction of the horizon of UHECR induced by energy losses more important at higher energies.

Once folded through the Galactic magnetic field, the dipole pattern expected from this mechanism is transformed into a more complex structure presumably described by a lower dipole amplitude and higher-order multipoles. However, in these scenarios, the dipole moment could remain the only one at reach within the sensitivity of the current generation of experiments. On the other hand, the detection of significant multipole moments beyond the dipole one could be suggestive of non-diffusive propagation of UHECRs from sources distributed in a non-isotropic way.

Future work will profit from the increased statistics, allowing us to uncover the angular distribution of UHECRs on the entire sky. This will provide further constraints helping to understand the origin of UHECRs.

## References

[1] The Pierre Auger and Telescope Array Collaborations, *ApJ* **794** (2014) 172.

[2] The Pierre Auger Collaboration, *ApJ* **802** (2015) 111.

[3] P. Sommers, *Astropart. Phys.* **14** (2001) 271.

[4] T. Abu Zayyad *et al.* (The Telescope Array Collaboration), *Astropart. Phys.* **48** (2013) 15.

[5] V. Verzi for the Pierre Auger Collaboration, Proc. 33rd ICRC, Rio de Janeiro, Brazil (2013) [arXiv:1307.5059].

[6] The Pierre Auger Collaboration, *JCAP* **08** (2014) 019.

[7] P. Billoir, O. Deligny, *JCAP* **02** (2008) 009.

[8] D. Harari, S. Mollerach, E. Roulet, *Phys. Rev. D* **89** (2014) 123001.

[9] P. Tinyakov, F. Urban, *J. Exp. Theor. Phys.* **120** 3 (2015) 533.

# Initial results of a direct comparison between the Surface Detectors of the Pierre Auger Observatory and the Telescope Array

**R. Takeishi[*a1], R. Cady[b], J.N. Matthews[b], T. Nonaka[a], S. Ogio[c], H. Sagawa[a] for the Telescope Array Collaboration[2]; C. Covault[d], T. Fujii[e], J. Johnsen[f], P. Lebrun[g], R. Lorek[d], P. Mantsch[g], J.A.J. Matthews[h], P. Mazur[g], S. Quinn[d], F. Sarazin[f], R. Sato[i], for the Pierre Auger Collaboration[3]; S. Collonges[j], B. Courty[j], B. Genolini[k], L. Guglielmi[j], M. Marton[l], E. Rauly[k], T. Trung[k], L. Smith[d], O. Wolf[f]**

[a] *Institute for Cosmic Ray Research, University of Tokyo, Kashiwa, Chiba, Japan*
[b] *High Energy Astrophysics Institute and Department of Physics and Astronomy, University of Utah, Salt Lake City, UT, USA*
[c] *Graduate School of Science, Osaka City University, Osaka, Osaka, Japan*
[d] *Department of Physics, Case Western Reserve University, Cleveland OH, USA*
[e] *University of Chicago, Enrico Fermi Institute, Chicago IL, USA*
[f] *Physics Department, Colorado School of Mines, Golden CO, USA*
[g] *Fermi National Laboratory, Batavia IL, USA*
[h] *Department of Physics and Astronomy, University of New Mexico, Albuquerque, NM, USA*
[i] *Pierre Auger Observatory, Malargüe, Argentina*
[j] *Laboratoire Astroparticules et Cosmologie (APC), Université Paris 7, CNRS-IN2P3, Paris, France*
[k] *Institut de Physique Nucléaire d'Orsay (IPNO), Université Paris 11, CNRS-IN2P3, Orsay, France*
[l] *Laboratoire de Physique Subatomique et de Cosmologie (LPSC), Université Joseph Fourier, INPG, CNRS-IN2P3, Grenoble, France*
[1] *E-mail:* take@icrr.u-tokyo.ac.jp
[2] *Full author list:* http://www.telescopearray.org/images/papers/ICRC2015-authorlist.pdf
[3] *Full author list:* http://www.auger.org/archive/authors_2015_ICRC.html

**Abstract:**
The Pierre Auger Observatory (Auger) in Mendoza, Argentina and the Telescope Array (TA) in Utah, USA aim at unraveling the origin and nature of Ultra-High Energy Cosmic Rays (UHECR). At present, there appear to be subtle differences between Auger and TA results and interpretations. Joint working groups have been established and have already reported preliminary findings. From an experimental standpoint, the Surface Detectors (SD) of both experiments make use of different detection processes not equally sensitive to the components of the extensive air showers making it to the ground. In particular, the muonic component of the shower measured at ground level can be traced back to the primary composition, which is critical for understanding the origin of UHECRs. In order to make direct comparisons between the SD detection techniques used by Auger and TA, a joint SD experimental research program is being developed. In the first phase, two Auger SD stations were deployed at the TA Central Laser Facility to compare station-level responses. This paper concentrates on the results obtained with the first Auger SD station (an "Auger North" design), which has been operating since October 2014. The second Auger SD station, identical to the ones being operated at Auger in Argentina (an "Auger South" design), was just deployed in June 2015. The second phase of this research program will be to co-locate six Auger North SD stations with TA stations in the field to compare event-level responses.

*Speaker.

## 1. Introduction

The Pierre Auger Observatory [1], which is located in Mendoza province, Argentina, is a hybrid instrument to detect cosmic-ray induced air showers. It combines a surface detector array (SD) with fluorescence telescopes making up the fluorescence detector (FD) overlooking the SD. The Auger SD consists of 1660 water-Cherenkov stations placed on a triangular grid with 1.5 km spacing, covering an area of 3000 $km^2$. The Telescope Array (TA) experiment [2], which is located in western Utah, USA, consists of 507 scintillation counters, placed on a square grid with 1.2 km spacing, covering 700 $km^2$, also combined with an array of FD telescopes. Both experiments study the spectrum, the origin, and the composition of UHECR.

Recently, joint working groups have been investigating some differences between Auger and TA results, e.g. the energy at which the flux suppression occurs [3], the interpretations of that suppression, and the composition of the UHECR primaries at the highest energies [4]. From a detection standpoint, the Auger and TA SD are not equally sensitive to the different air shower components. The Auger SD is sensitive to both the muonic and electromagnetic components and responds differently to each. The TA SD only measures charged particles and sees them all equally. Since most particles close to the core are electromagnetic, this is mostly what is measured. The muon component from air showers is an indicator of the primary cosmic ray composition and can provide a clue to revealing the origin and the acceleration mechanisms of UHECR. However, the number of muons observed with the Auger SD for each shower indicates a muon deficit in the air shower Monte-Carlo simulations [5]. The deficit ratio depends on the hadron interaction model. The number of muons observed in the data is 30% (EPOS LHC) to 80% (QGSJET-II-03) larger than that of the simulation assuming proton primaries at $10^{19}$ eV [6]. Understanding the origin of the discrepancy between measurements and models is critical.

In order to make direct comparisons between the SD detection techniques used by Auger and TA, a two-phase joint experimental research program is followed. The first phase consists in comparing station-level responses. In late October 2014, a water-Cherenkov tank (of the "Auger North" design [10]) was deployed at the TA Central Laser Facility (CLF) [7], where the TA muon detector project is also ongoing [8]. In June 2015, a second water-Cherenkov detector identical to those in Auger South was deployed adjacent to the Auger North SD station. The second phase will consist of deploying six Auger North SD stations alongside existing TA SD stations to allow for event-level comparisons of relatively low-energy showers with energies in the $10^{18}$ eV range. In this paper, we present the status and prospects of this joint research project, including the measurement of the first Auger North SD data that were recorded in coincidence with TA SD shower triggers.

## 2. Experiment

The Auger South SD station is a water-Cherenkov tank with three photomultiplier tubes (PMTs) that are symmetrically distributed at a distance of 1.2 m from the center of the tank [9]. The tank is 1.5 m tall and has a footprint of 10 $m^2$. It is filled to a depth of about 1.2 m with about 10 $M\Omega$ cm resistivity water. The water is contained in a flexible, laminated liner conforming approximately to the inner tank surface. The innermost lamination consists of Tyvek®. The Cherenkov light from air shower particles is diffusively reflected inside the water volume and viewed by the PMTs through optical windows. The signals are processed using front-end electronics having six 10-bit Fast Analog to Digital Converters (FADCs) running at 40 MHz. A dynamic range of 15 bits is realized using signals derived from the anode and from the last dynode ($\times 32$). The digitized signals are sent to a programmable logic device board used to make various triggering decisions.

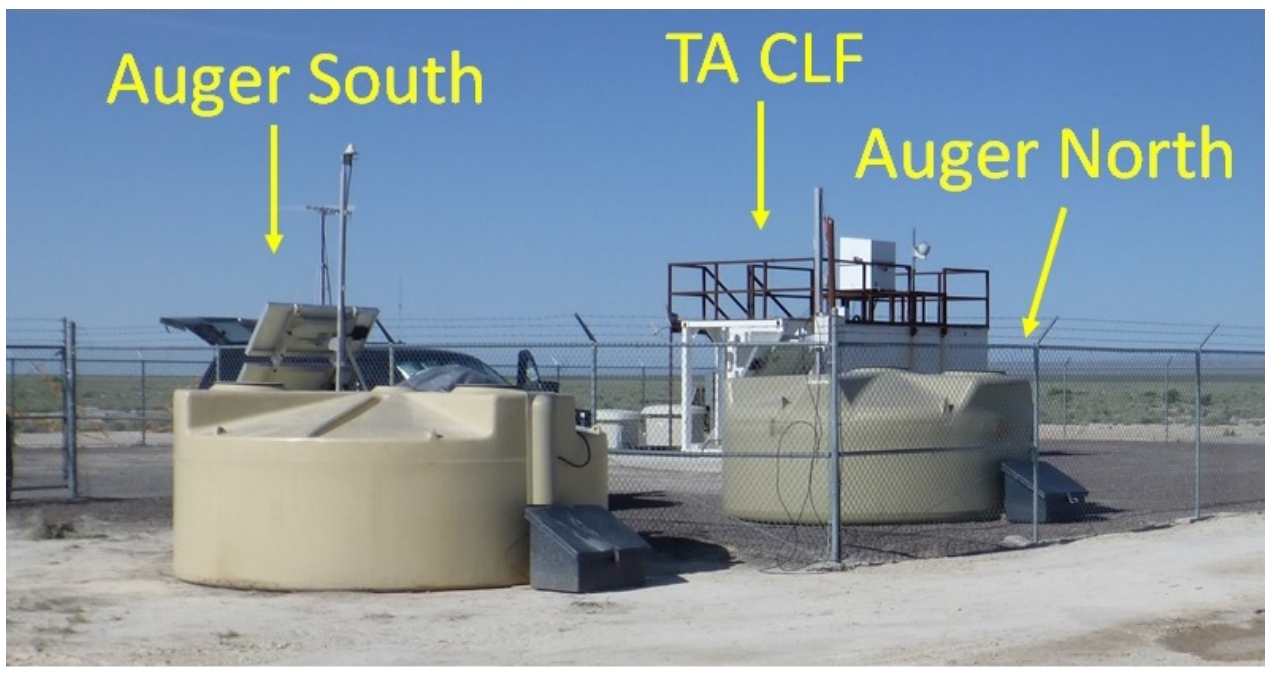

**Figure 1:** The two Auger SD stations deployed at the TA Central Laser Facility.

The Auger North SD station is a one PMT water Cherenkov surface detector used in the Pierre Auger Research and Development Array in Colorado, USA [11]. It is a cost-effective version of the Auger South SD station, with the same footprint, height and water volume. The Auger North and South SD stations deployed at the TA CLF are shown in figure 1. The design of the electronics for the Auger North surface detector is based on the one used at the Auger South SD. In this case however, the digitization is performed with commercial 10-bit ADCs with 100 MHz sampling rate. The dynamic range is extended to 22 bits, using signals derived from the anode ($\times 0.1$, $\times 1$ and $\times 30$) and from a deep (5th out of 8) dynode.

The TA SD station is composed of two layers of plastic scintillator with two PMTs, one for each layer [12]. It has an area of 3 $m^2$ and each layer has 1.2 cm thickness. The scintillators and PMTs are contained in a stainless steel box which is mounted under a 1.2 mm thick iron roof to protect the detector from large temperature variations. Photons that are generated in the scintillator are collected by wavelength shifting fibers and read out by PMTs. The signals from PMTs are digitized by a commercial 12-bit FADC with a 50 MHz sampling rate on the CPU board.

## 3. Analysis and Results

In order to start collecting data immediately after its deployment, the Auger North SD station was configured to record data locally. This was done by installing a large capacity (512GB) flash drive directly onto the local station controller. The second level trigger (T2) data, obtained from the standard Auger calibration procedure [1], were obtained and written on the drive at a rate of about 20 Hz. Only a very small fraction of those events arises from UHECR showers. A smaller dataset of atmospheric muons from the T1 trigger (100 Hz) was also collected to derive the Vertical Equivalent Muon (VEM) calibration from the single muon energy loss spectrum. In this analysis, the data from two observation periods are used; the first is Oct. 21, 2014 – Nov. 17, 2014 and the second is Nov. 19, 2014 – Dec. 7, 2014. The flash drive was swapped between the two periods. The exchange requires a shutdown of the station to open the tank and access the local station controller. Figure 2 (left) shows a typical waveform of the calibration data. Figure 2 (right) shows the histogram of pulse areas obtained from the FADC traces from the data and from the simulation. The air shower simulations were performed with CORSIKA in which the QGSJET-II-04 hadronic interaction model was used and the SD station geometry and response were simulated with GEANT4. The features of the spectrum are typical. The first peak corresponds to the residual noise suppressed by setting a threshold on the FADC integrated pulses. The second peak is the so-called “muon hump” from which the VEM can be deduced. As can be seen, the overall shape of the spectrum can be well described by the simulation. Figure 3 shows the evolution over time of the peak position of the single muon histogram. The VEM calibration appears to be relatively stable with small day/night (temperature-related) variation. An anomalous shift is observed around the time the flash drive is swapped. It is unclear at this time what caused this anomaly.

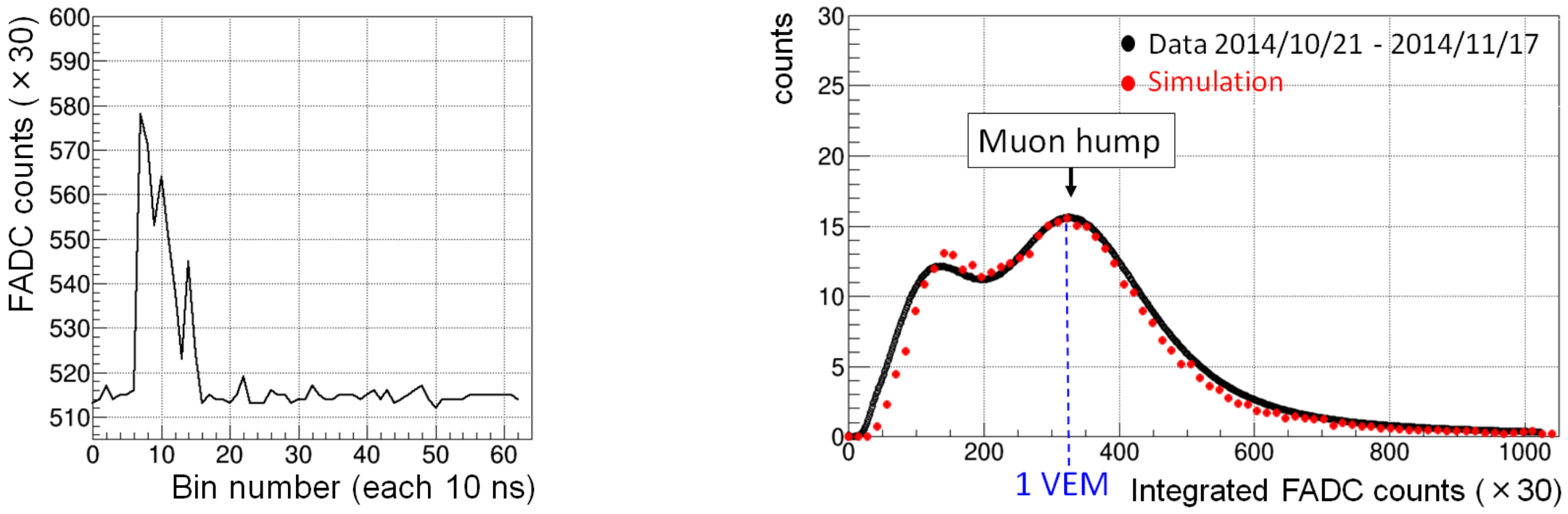

**Figure 2:** (left) A typical waveform of the calibration data of the Auger North SD. The anode ×30 signal is used. (right) The 1 VEM histogram from the calibration data and the MC simulation. The simulation is from a CORSIKA air shower simulation in which the QGSJET-II-04 hadronic interaction model and the Geant4 detector simulation are used. The peak position of the histogram for the simulation is adjusted to fit the experimental data.

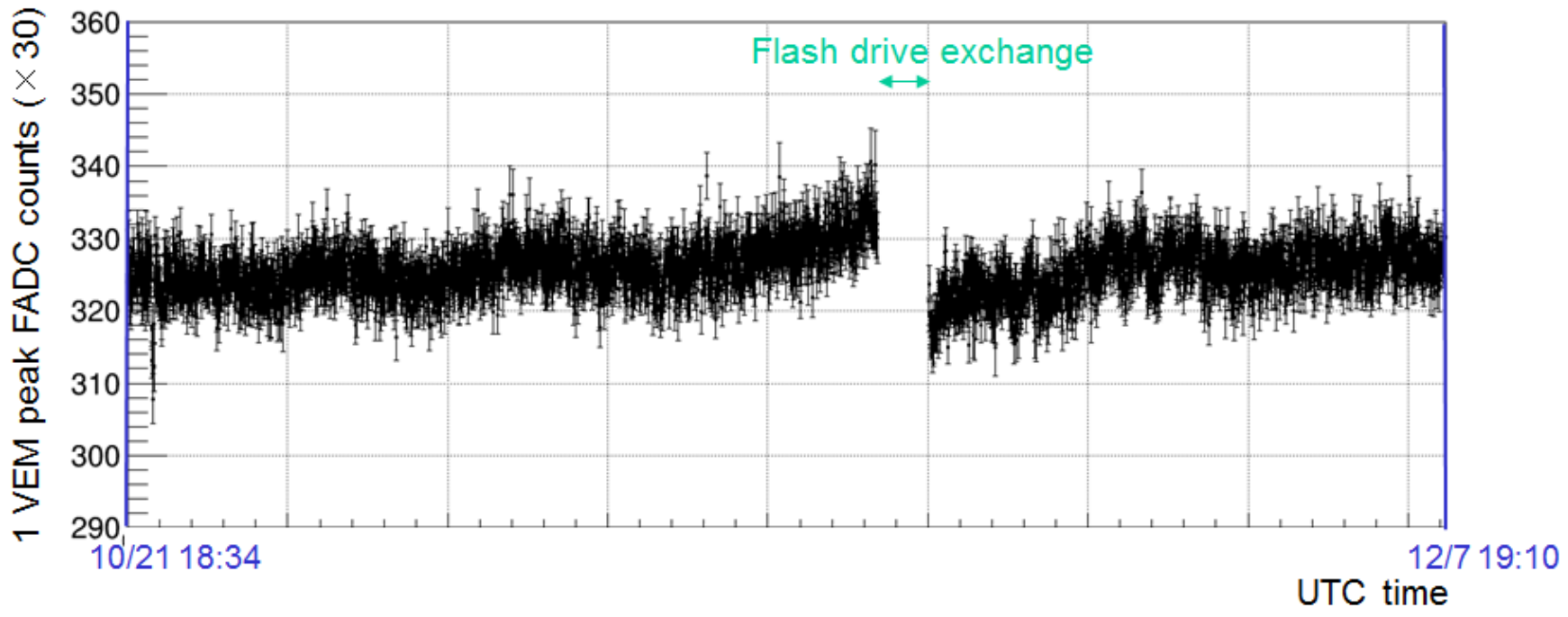

**Figure 3:** Time evolution of the peak position of the single muon histogram. Indicated on the figure is the time when the flash drive was swapped.

To identify actual UHECR shower events recorded by the Auger North SD station, the T2 timestamps are extracted and checked against the TA SD event trigger time within a $\pm 32\,\mu$s win-dow, corresponding to the time scale of a TA SD shower event. Air showers are reconstructed with the TA SD array using the method developed for the measurement of the energy spectrum by TA [16]. The air shower reconstruction program eliminates the events with poor resolution. In this analysis, we use the loose-cut selection criteria described in [17] to gain larger statistics.

Once the timestamp matching process is completed, 17 Auger North SD events are obtained in coincidence with TA SD loose-cut pass events. Among them, 10 events have core positions within 5 km of the Auger North SD. Figure 4 shows the display of a coincidence event. Also shown on the figure are the waveforms of the 5 TA SD and the Auger North SD stations associated with the air shower event. Based on the TA reconstruction, the energy and zenith angle of the primary cosmic ray are respectively $3.04\times10^{18}$ eV and 46.8°. Figure 5 (left) shows the waveform of the Auger North SD. The signal area is about 40 VEM ($\simeq$4 VEM/m$^2$) at about 1600 m from the shower core. The current event selection criterion for Auger-TA coincident events is that the Auger North SD trigger time is within $\pm 32\,\mu$s of the TA SD shower trigger time, but this condition possibly includes SDs triggered by background muons. Figure 5 (right) shows the corresponding distribution of the station trigger time as function of the distance between the stations and the shower core along the shower axis projected onto the ground. The Auger SD appears as the earliest trigger time among the SDs. The feature is consistent with the reconstructed shower geometry shown in figure 4. The other events present similar features.

Since March 2015, a third period of observation is ongoing, hence more results may be shown by the time of the conference. In June 2015, the second Auger SD station was deployed in the field. Both Auger SD stations at the CLF are now connected to a single board computer (SBC), which can be accessed remotely. In the near future, a TA SD station will be installed at the CLF and a local trigger will be formed between the Auger and TA SD stations, including the larger muon detector also installed at the CLF. Both the local and TA shower triggers will be provided to the Auger SBC to request data collection from the Auger SD stations. In this way, only data from higher level triggers will be collected. By combining the information of the three detector types, one will be able to thoroughly compare the response of the individual Auger and TA SD stations and estimate the relative contributions of the muon and EM components at the CLF location for showers triggered by TA.

The second phase of this joint collaboration aiming at making event-level comparisons will require the deployment of six contiguous Auger North SD stations alongside existing TA SD stations. The site for such a micro-Auger array has been identified and the deployment will occur towards the end of the year. The current plan is to have the Auger SD micro-array trigger independently to compare trigger efficiencies and energy estimators. Information from the Auger and TA lateral distribution functions should also provide insights on the muon and EM components of the observed showers.

## 4. Summary

A joint Auger-TA experimental research program studying the difference in SD responses is ongoing. Initial results are promising. The Auger North SD station VEM calibration appears to be well understood by the simulation, and first Auger-TA SD coincidences have been observed by matching the Auger SD station timestamps to the TA shower trigger in post processing. In June 2015, a second Auger (South) station was deployed in the field, and both Auger SD stations were connected to a SBC inside the CLF. This opens the door to the direct collection of higher-level local and global triggers allowing for more robust studies. The joint collaboration is also actively preparing the development of a micro-Auger array within TA for event-level comparative studies.

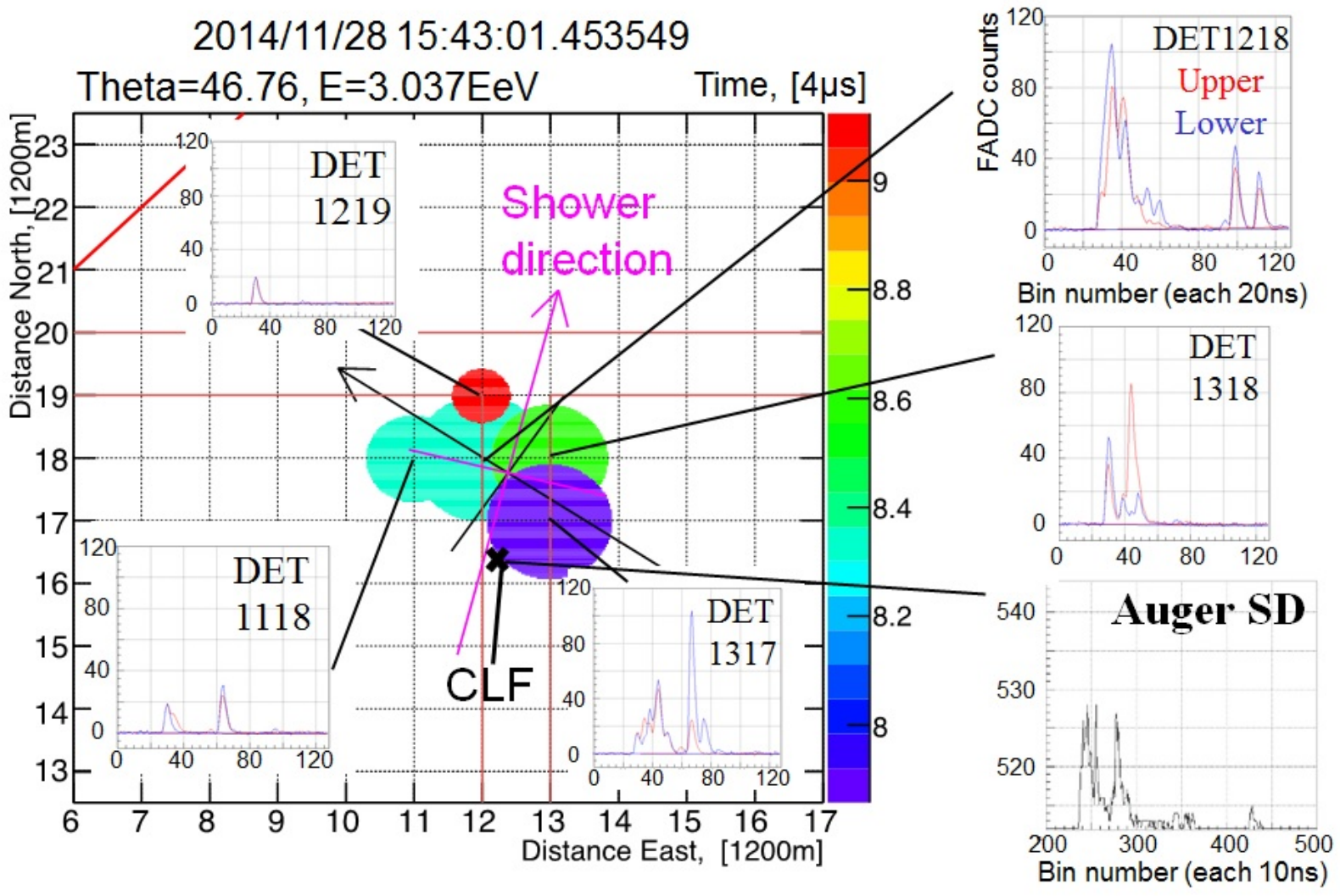

**Figure 4:** The event display for the event discussed in the text. The circle centers correspond to the TA SD positions. The circle areas are proportional to the logarithm of the TA SD station pulse heights. The circle colors represents the TA SD detection times from the detection time of the first triggered detector within the same $\pm$ 32 $\mu$s. The magenta arrow shows the shower axis projected onto the ground and the black x mark is the position of the Auger North SD at the CLF. The red line of the TA SD waveform is the upper layer signal, while the blue line shows the lower layer signal.

## References

[1] A. Aab *et al.* (Pierre Auger Collaboration), *The Pierre Auger Cosmic Ray Observatory*, *Nucl. Instrum. Meth. A* **798** (2015) 172–213; arXiv:1502.01323.

[2] P. Tinyakov *et al.* (Telescope Array Collaboration), *Latest results from the telescope array*, *Nucl. Instrum. Meth. A* **742** (2014) 29–34.

[3] A. Aab *et al.*, (Pierre Auger collaboration), *Measurement of the cosmic ray spectrum above* $4{\times}10^{18}$ *eV using inclined events detected with the Pierre Auger Observatory*, submitted to *JCAP* (2015) arXiv:1503.07786.

[4] R. Abbasi *et al.*, for the Pierre Auger Collaboration and the Telescope Array Collaboration, *Report of the Working Group on the Composition of Ultra High Energy Cosmic Rays*, to appear in the Proceedings of the UHECR workshop, Springdale, USA (2014) arXiv:1503.07540.

[5] A. Yushkov for the Pierre Auger Collaboration, *Measurement of the muon shower content at the Pierre Auger Observatory*, *EPJ. Conf.* **53** (2013) 07002.

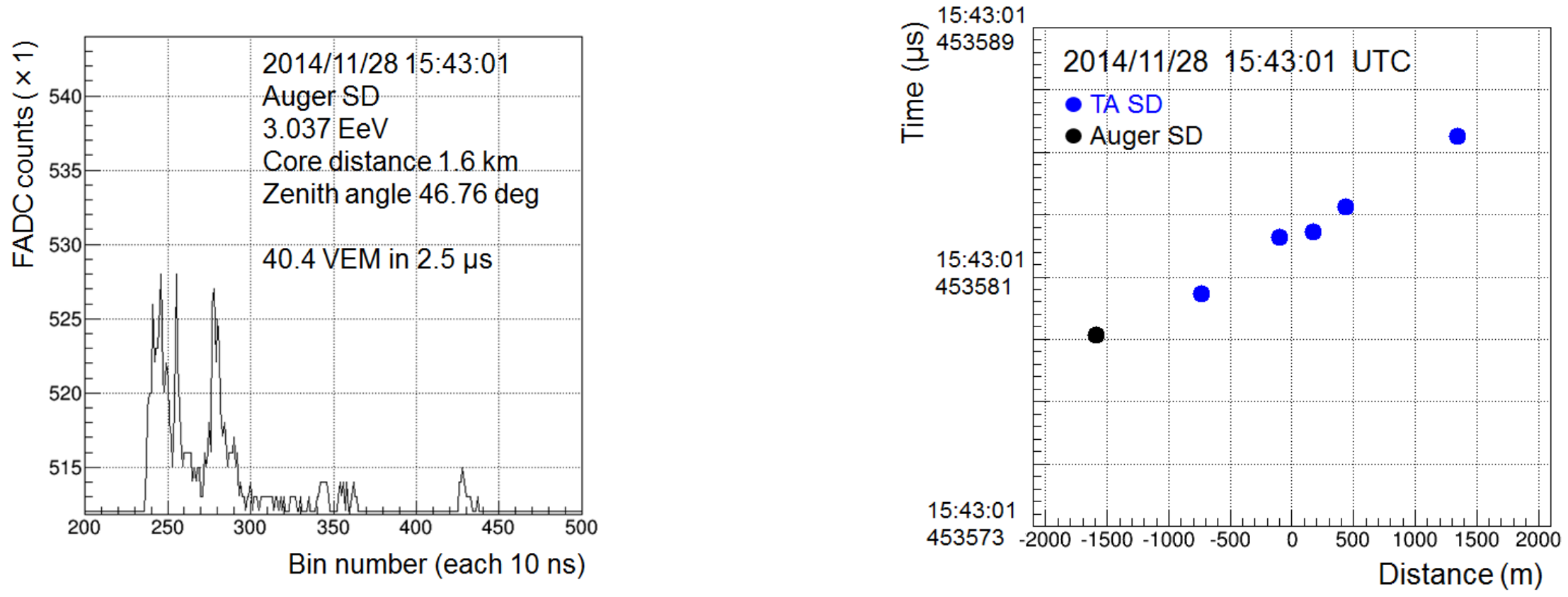

**Figure 5:** (left) The waveform of the Auger North SD station for the event discussed in the text. (right) The distribution of the station trigger time as function of the distance between the stations and the shower core along the shower axis projected onto the ground.

[6] A. Aab *et al.*, (The Pierre Auger Collaboration), *Muons in air showers at the Pierre Auger Observatory: Mean number in highly inclined events*, *Phys. Rev. D.* **91** (2015) 032003, arXiv:1408.1421.

[7] S. Udo *et al.*, *The Central Laser Facility at the Telescope Array*, *Proc. 30th ICRC* **5** (2007) 1021, Merida, Mexico.

[8] T. Nonaka *et al.*, *Performance and status of TA muon detector*, these proceedings.

[9] I. Allekotte *et al.*, *The surface detector system of the Pierre Auger Observatory*, *Nucl. Instrum. Meth. A* **586** (2008) 409–420, arXiv:0712.2832.

[10] J. Blümer *et al.* (The Pierre Auger Collaboration), *The northern site of the Pierre Auger Observatory*, *New J. Phys.* **12** (2010) 035001.

[11] F. Sarazin *et al.*, *The Pierre Auger Research and Development Array (RDA) in southeastern Colorado – R&D for a giant ground array*, *EPJ. Conf.* **53** (2013) 08017.

[12] T. Abu-Zayyad *et al.*, *The surface detector array of the Telescope Array experiment*, *Nucl. Instrum. Meth. A* **689** (2012) 87–97.

[13] M. Teshima *et al.*, *Properties of $10^9$–$10^{10}$ GeV extensive air showers at core distances between 100 and 3000 m*, *J. Phys. G* **12** (1986) 1097–1113.

[14] M. Takeda *et al.*, *Extension of the Cosmic-Ray Energy Spectrum beyond the Predicted Greisen-Zatsepin-Kuz'min Cutoff*, *Phys. Rev. Lett.* **81** (1998) 1163–1166.

[15] M. Takeda *et al.*, *Energy determination in the Akeno Giant Air Shower Array experiment*, *Astropart. Phys.* **19** (2003) 447–662.

[16] T. Abu-Zayyad *et al.*, *The Cosmic Ray Energy Spectrum Observed with the Surface Detector of the Telescope Array Experiment*, *ApJ* **768**(2013) L1.

[17] R.U. Abbasi *et al.*, *Indications of Intermediate-Scale Anisotropy of Cosmic Rays with Energy Greater than 57 EeV in the Northern Sky Measured with the Surface Detector of the Telescope Array Experiment*, *ApJ* **790** (2014) L21.